\documentclass[preprint,pteplogo]{ptephy_v2}

\preprintnumber{XXXX-XXXX}

\usepackage{hyperref}  
\usepackage{lineno}
\usepackage[percent]{overpic}
\usepackage{acro}
\usepackage{xspace}
\usepackage{academicons} 
\newcommand{\orcid}[1]{\href{https://orcid.org/#1}{\textcolor[HTML]{A6CE39}{\aiOrcid}}}

\DeclareAcronym{submet}{
  short = SUBMET,
  long  = SUB-Millicharge ExperimenT,
  tag = abbrev
}

\DeclareAcronym{lhc}{
  short = LHC,
  long  = Large Hadron Collider,
  tag = abbrev
}

\DeclareAcronym{pmt}{
  short = PMT,
  long  = photomultiplier tube,
  tag = abbrev
}

\DeclareAcronym{t2k}{
  short = T2K,
  long  = Tokai to Kamioka Experiment,
  tag = abbrev
}

\DeclareAcronym{daq}{
  short = DAQ,
  long  = data acquisition,
  tag = abbrev
}

\DeclareAcronym{rms}{
  short = RMS,
  long  = root mean square,
  tag = abbrev
}

\DeclareAcronym{spe}{
  short = SPE,
  long  = single-photoelectron,
  tag = abbrev
}

\DeclareAcronym{led}{
  short = LED,
  long  = light-emitting diode,
  tag = abbrev
}

\DeclareAcronym{ttl}{
  short = TTL,
  long  = transistor-transistor logic,
  tag = abbrev
}

\DeclareAcronym{fem}{
  short = FEM,
  long  = finite element method,
  tag = abbrev
}

\DeclareAcronym{pcb}{
  short = PCB,
  long  = printed circuit board,
  tag = abbrev
}

\DeclareAcronym{jparc}{
  short = J-PARC,
  long  = Japan Proton Accelerator Research Complex,
  tag = abbrev
}

\usepackage{bm,epstopdf,natbib,hyperref,color,verbatim,multirow,bm,tikz,xcolor}
\hypersetup{colorlinks=true,urlcolor=blue,citecolor=blue,linkcolor=blue,menucolor=blue,anchorcolor=blue,filecolor=blue}
\everymath{\displaystyle}
\definecolor{lime}{HTML}{A6CE39}
\DeclareRobustCommand{\orcidicon}{
  \begin{tikzpicture}
    \draw[lime, fill=lime] (0,0) 
    circle [radius=0.16] 
    node[white] {{\fontfamily{qag}\selectfont \tiny ID}};
    \draw[white, fill=white] (-0.0625,0.095) 
    circle [radius=0.007];
  \end{tikzpicture}
  \hspace{-3mm}
}
\foreach \x in {A, ..., Z}{\expandafter\xdef\csname orcid\x\endcsname{\noexpand\href{https://orcid.org/\csname orcidauthor\x\endcsname}
  {\noexpand\orcidicon}}
}

\begin{document}

\title{Design and Mechanical Integration of Scintillation Modules for SUB-Millicharge ExperimenT (SUBMET)}

\author[1]{Claudio Campagnari\hspace{-1mm}\orcidP{}}
\author[2]{Sungwoong Cho\thanks{Present address: Department of General Studies, Hongik University, 22639 Sejong-ro, Jochiwon-eup, Sejong, Korea}}
\author[2]{Suyong Choi\hspace{-1mm}\orcidG{}}
\author[2]{Seokju Chung\hspace{-1mm}\orcidH{}\thanks{Present address: Columbia University, 116th and Broadway, New York, New York 10027, USA}}
\author[3]{Matthew Citron\hspace{-1mm}\orcidF{}}
\author[4]{Albert De Roeck\hspace{-1mm}\orcidI{}}
\author[4]{Martin Gastal}
\author[2]{Seungkyu Ha\hspace{-1mm}\orcidO{}}
\author[5]{Andy Haas}
\author[6]{Christopher Scott Hill\hspace{-1mm}\orcidJ{}}
\author[2]{Byeong Jin Hong}
\author[2]{Haeyun Hwang}
\author[2]{Insung Hwang\hspace{-1mm}\orcidD{}\thanks{Present address: Boston University, Commonwealth Ave, Boston, Massachusetts 02215, USA}}
\author[2]{Hoyong Jeong\hspace{-1mm}\orcidB{}}
\author[2]{Hyunki Moon\hspace{-1mm}\orcidC{}}
\author[2]{Jayashri Padmanaban\hspace{-1mm}\orcidK{}}
\author[1]{Ryan Schmitz\hspace{-1mm}\orcidE{}}
\author[2]{Changhyun Seo}
\author[1]{David Stuart\hspace{-1mm}\orcidL{}}
\author[2]{Eunil Won\hspace{-1mm}\orcidM{}}
\author[2]{Jae Hyeok Yoo\hspace{-1mm}\orcidA{}\thanks{Email: jaehyeokyoo@korea.ac.kr}}
\author[2]{Jinseok Yoo}
\author[7]{Ayman Youssef}
\author[7]{Ahmad Zaraket}
\author[7]{Haitham Zaraket\hspace{-1mm}\orcidN{}}

\affil[1]{Department of Physics, University of California, Santa Barbara, California 93106, USA}
\affil[2]{Department of Physics, Korea University, 145 Anam-ro, Seongbuk-gu, Seoul 02841, Korea }
\affil[3]{Department of Physics, University of California, One Shields Avenue, Davis, California 95616, USA}
\affil[4]{CERN, CH-1211 Geneva, Switzerland}
\affil[5]{Department of Physics, New York University,  726 Broadway, New York, New York 10012, USA}
\affil[6]{Department of Physics, The Ohio State University, 191 West Woodruff Ave, Columbus, Ohio 43210, USA}
\affil[7]{Multidisciplinary Physics Lab, Lebanese University, RGHC+4PR, Hadeth-Beirut, Lebanon}

\begin{abstract}
We present a detailed description of the detector design for the SUB-Millicharge ExperimenT (SUBMET), developed to search for millicharged particles. The experiment probes a largely unexplored region of the charge-mass parameter space, focusing on particles with mass $m_\chi < 1.6~\textrm{GeV}/c^2$ and electric charge $Q < 10^{-3}e$. The detector has been optimized to achieve high sensitivity to interactions of such particles while maintaining effective discrimination against background events. We provide a comprehensive overview of the key detector components, including scintillation modules, photomultiplier tubes, and the mechanical support structure. Furthermore, we present the results of weight and seismic analyses, which validate the structural integrity and operational safety of the detector under various conditions.
\end{abstract}

\subjectindex{C30, H10}
\maketitle

\section{Introduction}\label{sec:intro}
A wide range of astrophysical and cosmological evidence has suggested the existence of dark matter. However, the nature of this enigmatic substance remains unknown, and the direct detection of dark matter particles is still an open challenge in contemporary experimental physics. A broad class of theories postulates the existence of a dark sector that only weakly interacts with ordinary matter. In these theories, particles from the dark sector can serve as dark matter candidates. One intriguing possibility is that of millicharged particles ($\chi s$), which can arise as a low-energy consequence of a new, massless $\mathrm{U}(1)'$ gauge boson, a dark photon~\cite{HOLDOM1986196}.

The \ac{submet}~\cite{submet} is designed to search for millicharged particles at \ac{jparc}, which can be produced from the 30~GeV proton-target collisions. The detector consists of arrays of long scintillator bars, intended to increase the likelihood that millicharged particles will interact with the detector material. The scintillator bars are housed in an aluminum structure that is robust against both static and seismic stress.

The rest of the paper is organized as follows.
After an overview of the detector design in Section~\ref{sec:overview}, Section~\ref{sec:modules} describes the design and construction of the scintillator and \ac{pmt}s, which form the core of the detection mechanism. The mechanical support for the detector system is elaborated in Section~\ref{sec:mech_sup}. Section~\ref{sec:fea} provides an in-depth explanation of the mechanical support structure, with a focus on the weight and seismic analyses. The paper concludes with a summary and future perspectives for the \ac{submet} experiment.

\section{Overview of detector design}\label{sec:overview}
The basic idea of the experiment is to search for feebly interacting particles, such as millicharged particles $\chi$s, produced in high-energy proton collisions. At J-PARC, a 30 GeV proton beam is directed onto a graphite target, producing a wide range of mesons $\mathfrak{m}$s. Lighter mesons like $\pi^0$ and $\eta$ primarily decay through photons ($\mathfrak{m} \rightarrow \gamma \chi \bar{\chi}$), while heavier mesons such as $\rho$, $\omega$, $\phi$, and $J/\psi$ can decay directly into pairs of $\chi$s ($\mathfrak{m} \rightarrow \chi \bar{\chi}$). These $\chi$s are expected to propagate along or near the beam axis. The \ac{submet} detector is positioned 280~m downstream from the target and offset by about 1~degrees below the beam axis. As a result, the detector must be tilted upward by 4.5~degrees with respect to the floor to align with the expected $\chi$ flux. To maximize sensitivity, it is crucial to efficiently cover the angular region where $\chi$ particles are likely to travel. This motivates a detector design that maximizes the active volume within the available space, particularly in the direction of $\chi$ propagation. To achieve this, the \ac{submet} detector employs a segmented design based on long scintillator bars. A large sensitive volume is essential to reach the target sensitivity to electric charges as small as $10^{-3}e$ or below. Segmenting this volume offers several advantages: it significantly reduces background from \ac{pmt} dark current pulses and from cosmogenic muon-induced showers, and it enables directional reconstruction to suppress events inconsistent with an origin at the target.

\begin{figure}[h]
\begin{center}
\includegraphics[width=0.9\linewidth]{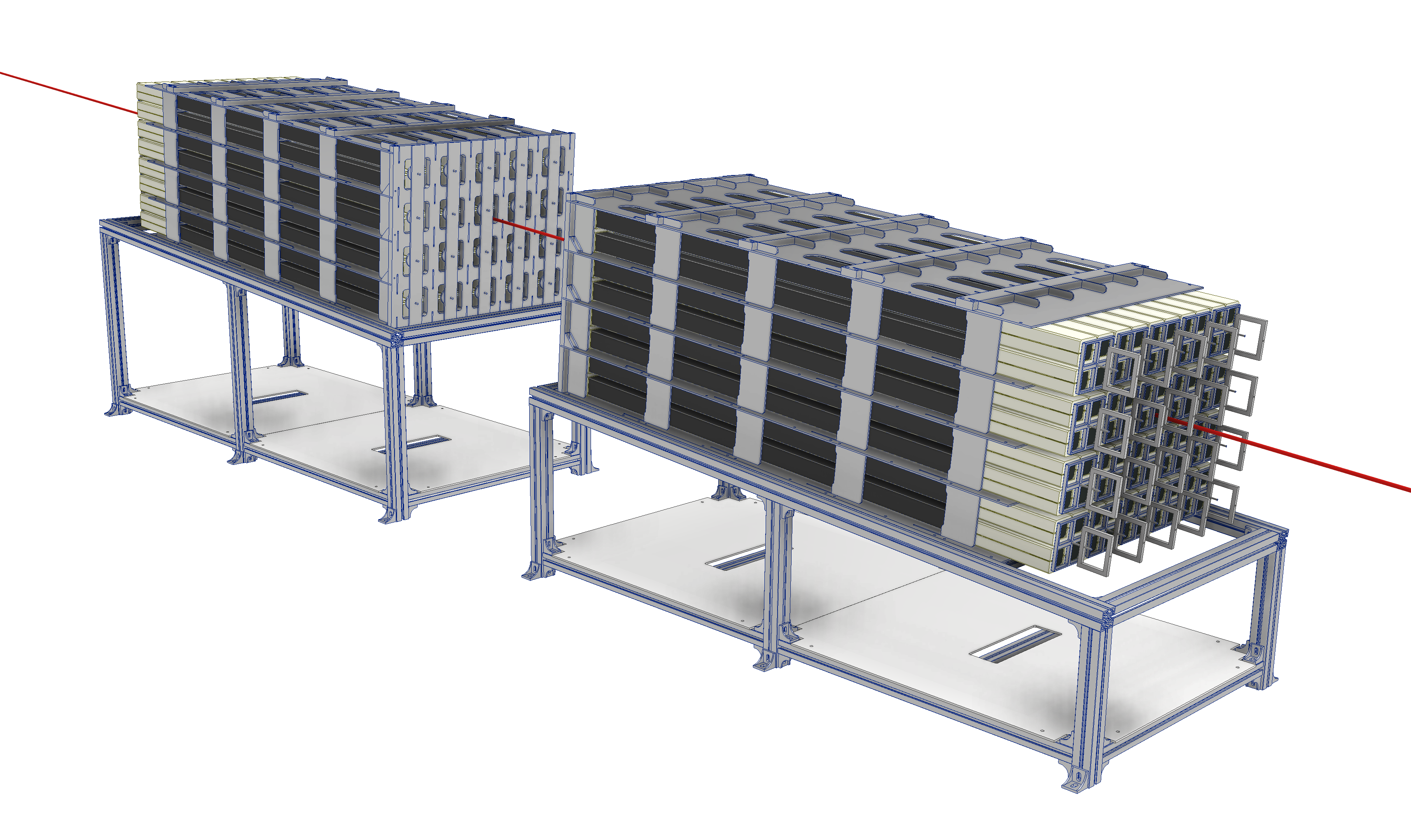}
\end{center}
\caption{A 3D model of the \ac{submet} detector. Each module consists of a $50 \times 50 \times 1500~\textrm{mm}^3$ plastic scintillator bar coupled to a \ac{pmt}. Four modules are grouped in a $2 \times 2$ configuration to form a supermodule. Each layer comprises $5 \times 4$ supermodules. The detector consists of two such layers, aligned to ensure that a millicharged particle, shown as a red line, passes through both layers from left to right within a narrow time window.}
\label{fig:detector}
\end{figure}
As illustrated in Figure~\ref{fig:detector}, the detector comprises two layers of stacked scintillator bars, arranged so that $\chi$s passing through the detector are likely to traverse both layers within a narrow coincidence time window. Each layer consists of $10 \times 8$ bars. A prototype with a similar design has been successfully tested at the \ac{lhc}, demonstrating both its robustness and its sensitivity to $\chi$ detection~\cite{PhysRevD.102.032002}.

Each scintillator bar has a \ac{pmt} attached at one end to convert photons into electronic signals. \ac{pmt}s are well suited for this application due to their large area coverage, low cost, and low dark current. A scintillator bar paired with a \ac{pmt} is referred to as a ``module.'' Four modules are grouped in a $2 \times 2$ configuration to form a ``supermodule,'' which simplifies installation and maintenance.

The supermodules are supported by aluminum frames with a shelf-like structure. Separate frames are used for each layer and are mounted onto a supporting frame made of aluminum profiles.

\section{Modules}\label{sec:modules}
A module refers to a single, independent detection unit consisting of a scintillator bar, a \ac{pmt}, and auxiliary support structures. This section describes the configuration, key characteristics, primary functions, and assembly process of a module.

A module is constructed by attaching a Hamamatsu Photonics R7725 \ac{pmt}~\cite{r7725} to one end of an EJ-200 plastic scintillator bar~\cite{ej200}. The scintillator bar produces light when a millicharged particle passes through it, while the \ac{pmt} converts the light into an electronic signal that can be recorded by the readout electronics. The auxiliary structures provide mechanical support and help maintain the alignment between the scintillator bar and the \ac{pmt}.

\subsection{Scintillator}\label{sec:scint}
Eljen Technology’s EJ-200 scintillator~\cite{ej200} offers a long attenuation length, fast scintillation timing, and cost-effectiveness, making it an ideal choice for constructing a large-scale detector. Its main properties are summarized in Table~\ref{tab:ej200}.
\begin{table}
\begin{center}
\caption{\label{tab:ej200} Main properties of EJ-200~\cite{ej200}.}
\begin{tabular}{c c} 
 \hline
 Properties & Values \\
 \hline\hline
 Scintillation efficiency (photons/1~MeV~$e^-$) & 10,000  \\
 Wavelength of Maximum Emission                 & 425~nm  \\
 Light Attenuation Length                       & 3800~mm \\ 
 Rise Time                                      & 0.9~ns  \\
 Decay Time                                     & 2.1~ns  \\
 \hline
\end{tabular}
\end{center}
\end{table}
The dimensions of a scintillator bar were chosen to be \(50 \times 50 \times 1500~\textrm{mm}^3\). The width of 50~mm was determined by the radius of the \ac{pmt}. A {\textsc{Geant4}\xspace}~\cite{geant4} simulation was performed to optimize the length. In the simulation, a 5~GeV muon was fired along the long axis of the scintillator, assuming a surface reflectivity of 98\%. The number of scintillation photons reaching one end of the bar (\(N_\gamma\)) was counted as the length was varied.

\begin{figure}[h]
\begin{center}
\includegraphics[width=0.7\linewidth]{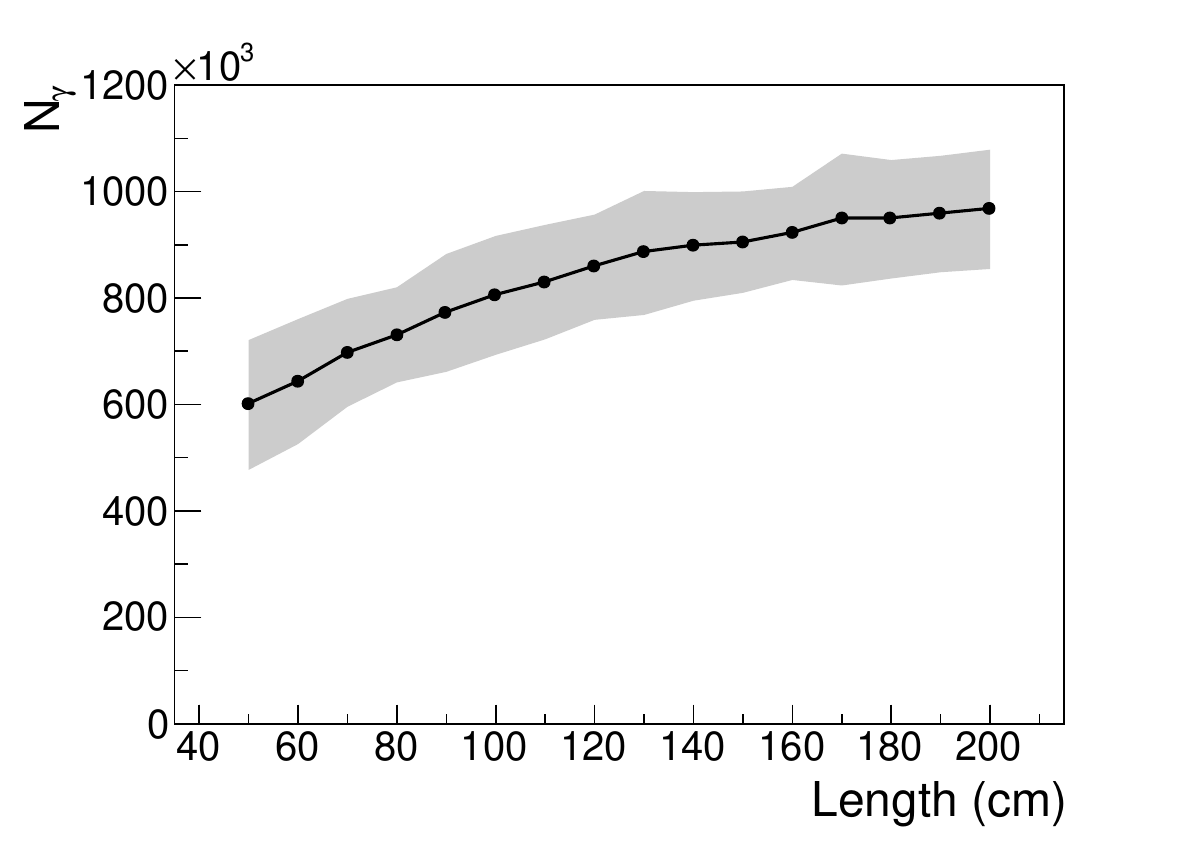}
\end{center}
\caption{The mean number of simulated scintillation photons reaching one end of the scintillator bar ($\bar{N}_\gamma$) as a function of bar length. The gray band represents the \ac{rms} for each length.}
\label{fig:bar_length_opt}
\end{figure}
Figure~\ref{fig:bar_length_opt} illustrates the mean number of scintillation photons, \(\bar{N}_\gamma\). Due to limited space at the experimental site, the maximum bar length is approximately 1500~mm. Since \(\bar{N}_\gamma\) changes only slightly beyond this length, the spatial constraint has minimal impact on performance of the detector.

Each scintillator bar is wrapped in two layers of 100~\textmu m-thick Teflon, one layer of 18~\textmu m-thick aluminum foil, and a 160~\textmu m-thick insulating tape. The aluminum foil, wrapped over the Teflon layers, blocks and reflects any residual light. Finally, the insulating tape is wrapped around the assembly to completely block ambient light. The addition of these three layers increases the overall thickness of the scintillator bar by approximately 1~mm.

\begin{figure}[h]
\centering
\begin{tabular}{cc}
  \begin{overpic}[width=0.4\textwidth]{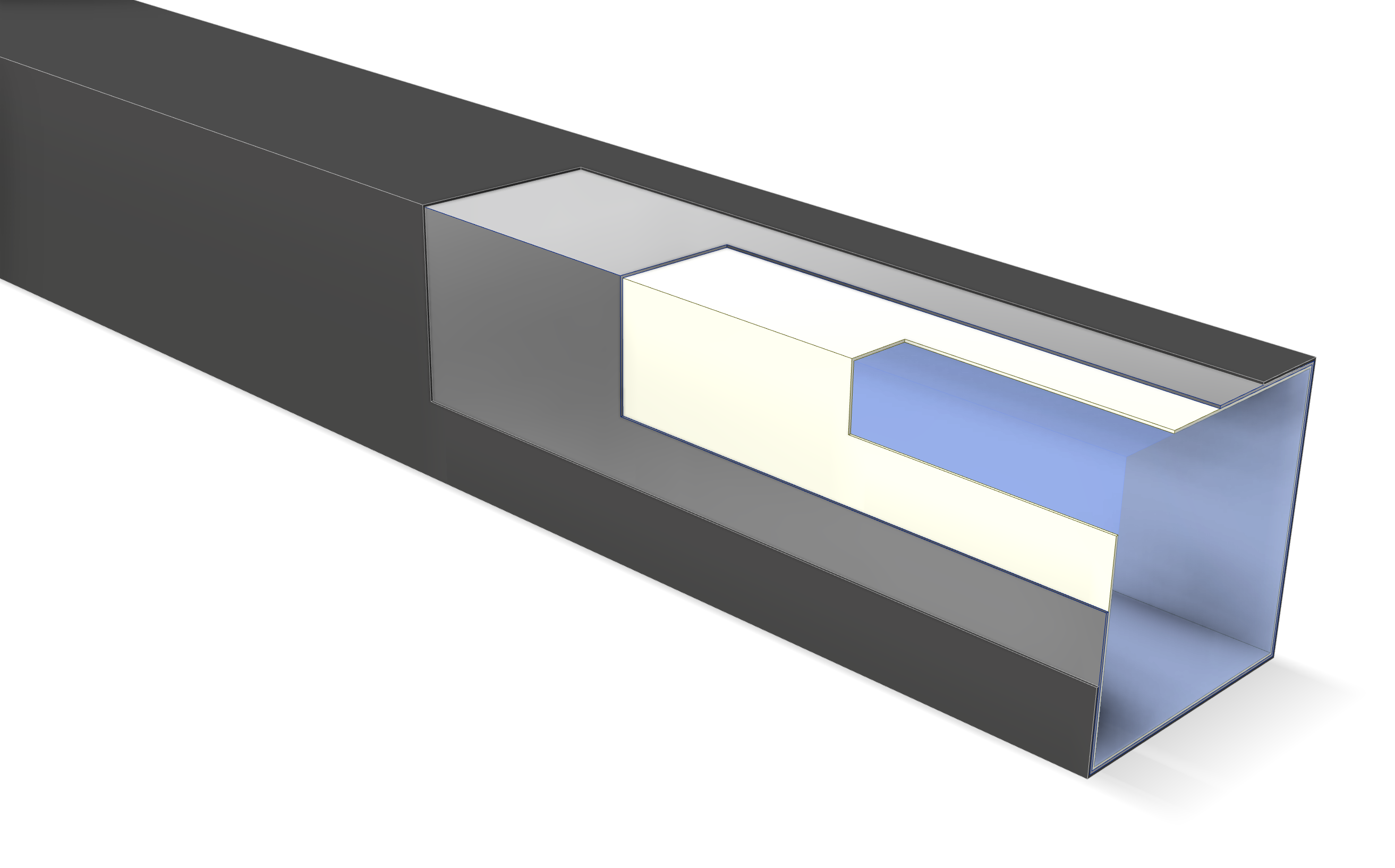}
    \put(85, 20){EJ-200}
    \put(50, 33){$2\times100$~\textmu m teflon}
    \put(48, 45){$2\times18$~\textmu m Al foil}
    \put(25, 60){$2\times160$~\textmu m insulating tape}
  \end{overpic} &
  \includegraphics[width=0.4\textwidth]{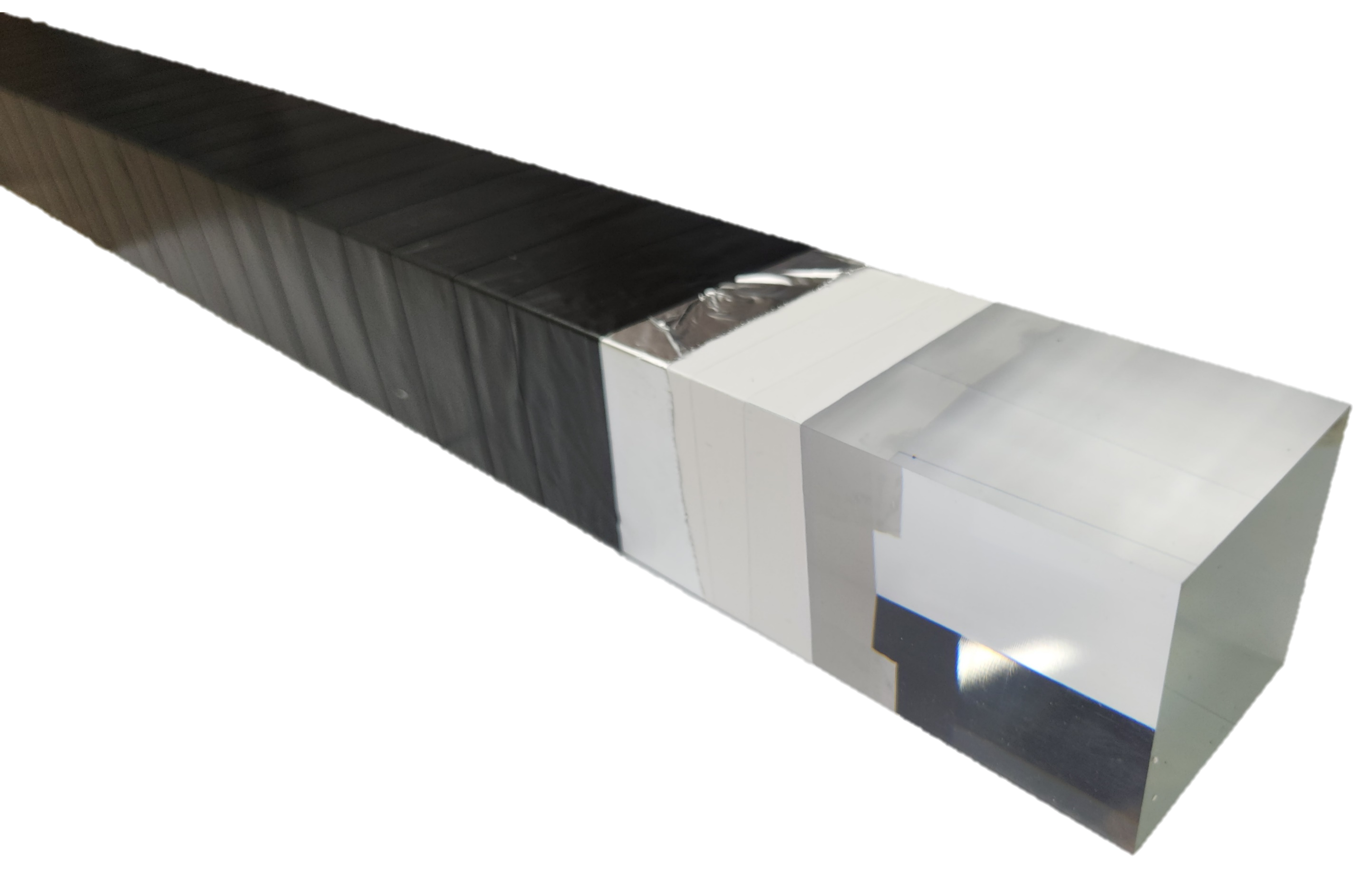} \\
  (a) & (b)
\end{tabular}
\caption{\label{fig:scint} (a) 3D model of the scintillator wrapping layers. (b) Photograph of a partially wrapped scintillator bar. The wrapping is intentionally left exposed to show the individual layers.}
\end{figure}
Figure~\ref{fig:scint} (a) and (b) depict the order of the wrapping and an image of a scintillator, respectively.

\subsection{Support}\label{sec:support}
\begin{figure}[ht]
\centering
\begin{tabular}{cc}
  \includegraphics[width=0.4\textwidth]{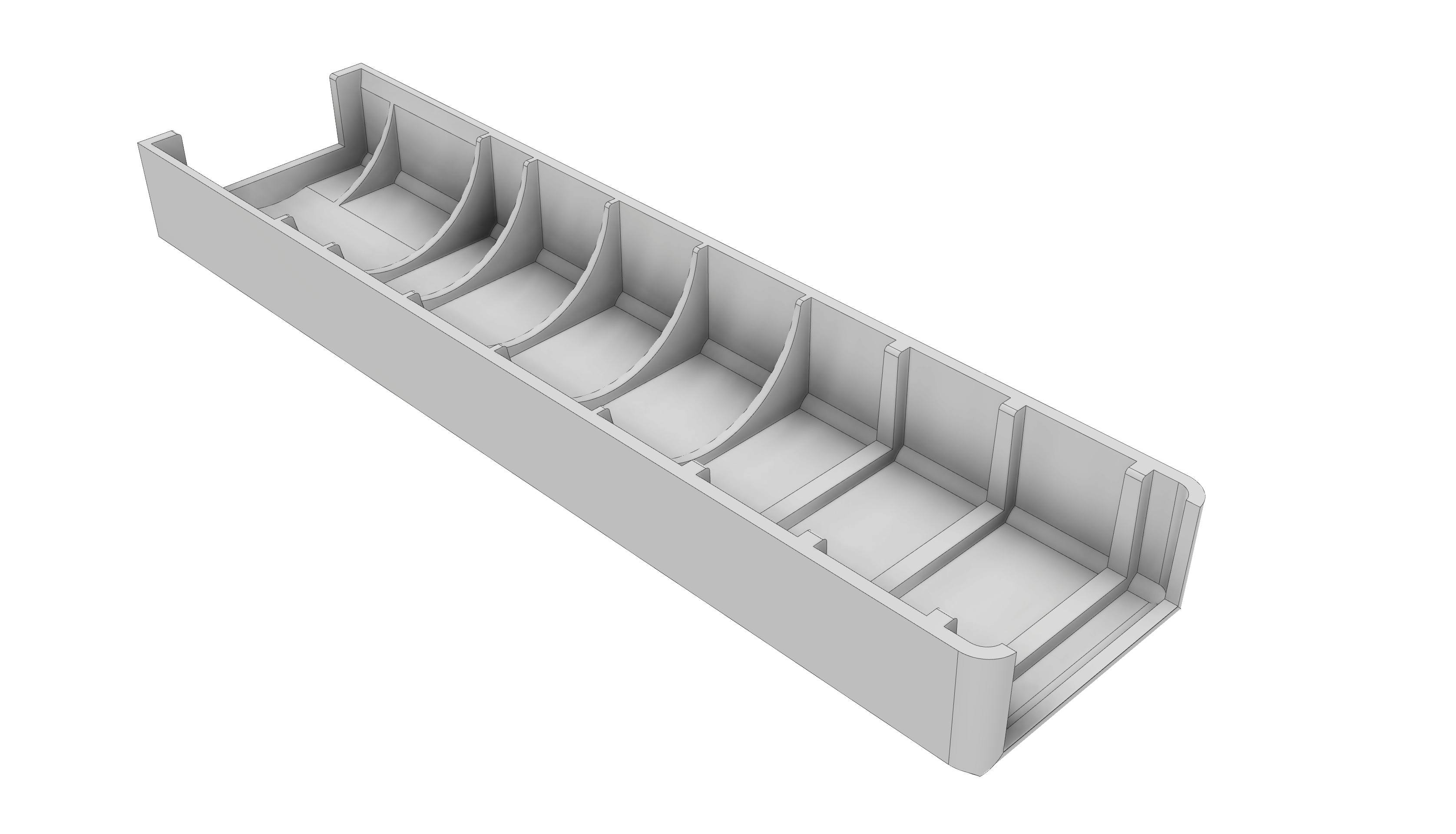} &
  \includegraphics[width=0.25\textwidth]{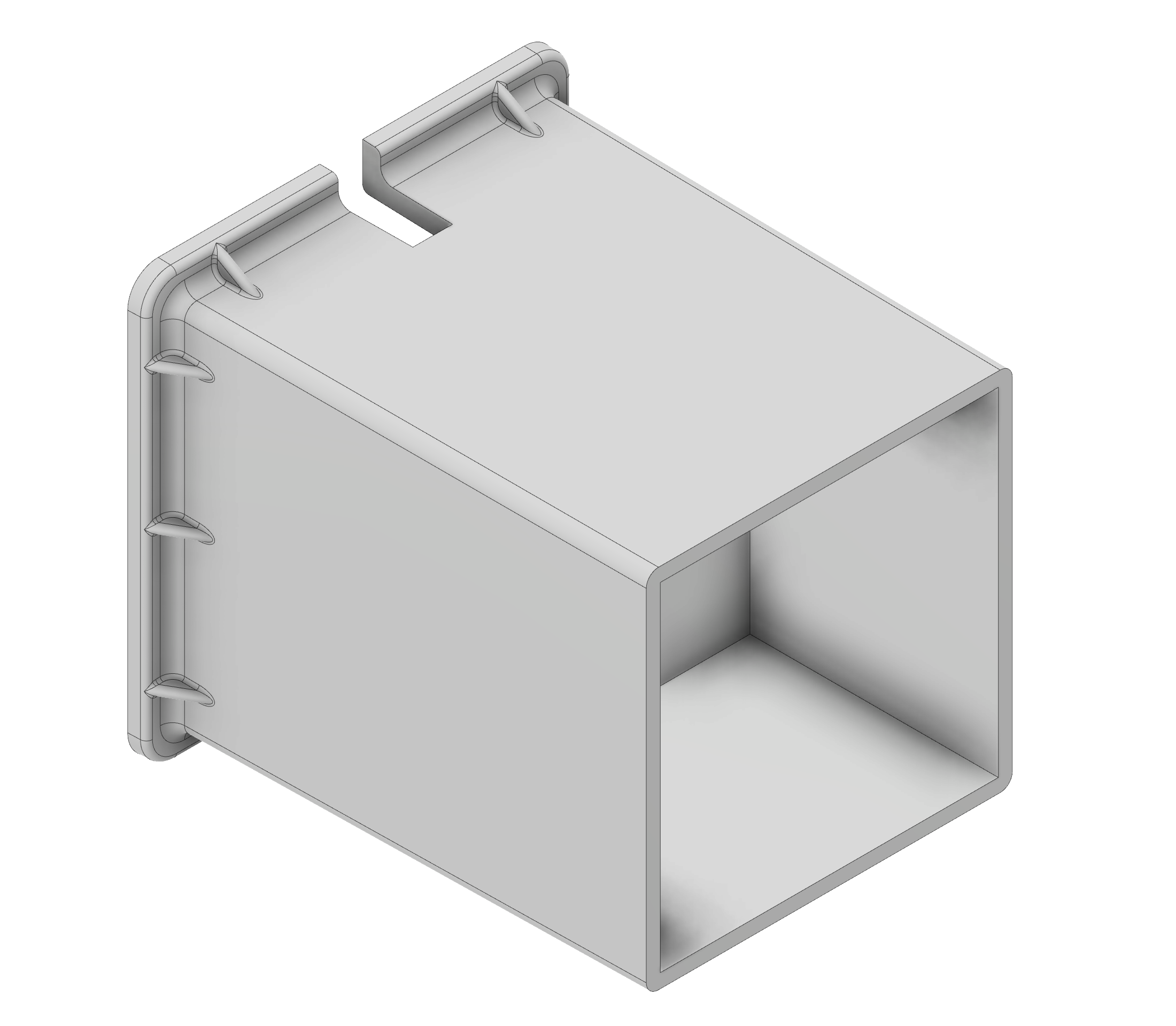} \\
  (a) & (b)
\end{tabular}
\caption{\label{fig:support} 3D models of (a) the ``\ac{pmt} Support'' and (b) the ``\ac{led} Support'' components. 
}
\end{figure}

A plastic support structure is designed to strengthen both the optical and mechanical coupling between a scintillator and a \ac{pmt}. It applies longitudinal pressure to ensure a firm connection, preventing the \ac{pmt} from detaching due to gravity. Figure~\ref{fig:support}(a) shows a 3D model of the design. Two pieces are used to fully enclose the \ac{pmt} and the corresponding end of the scintillator; this structure is referred to as the ``\ac{pmt} support.'' The assembly process is described in Section~\ref{sec:mod_assem}.

The ``\ac{led} support'' is attached to the opposite end of the scintillator, where the \ac{led} is located. It houses an \ac{led} circuit used for various testing purposes. Its lateral dimensions match those of the \ac{pmt} support to maintain a uniform spacing between adjacent scintillators.

\subsection{Module Assembly}\label{sec:mod_assem} 
\begin{figure}[h]
\centering
\begin{tabular}{cc}
  \includegraphics[width=0.4\textwidth]{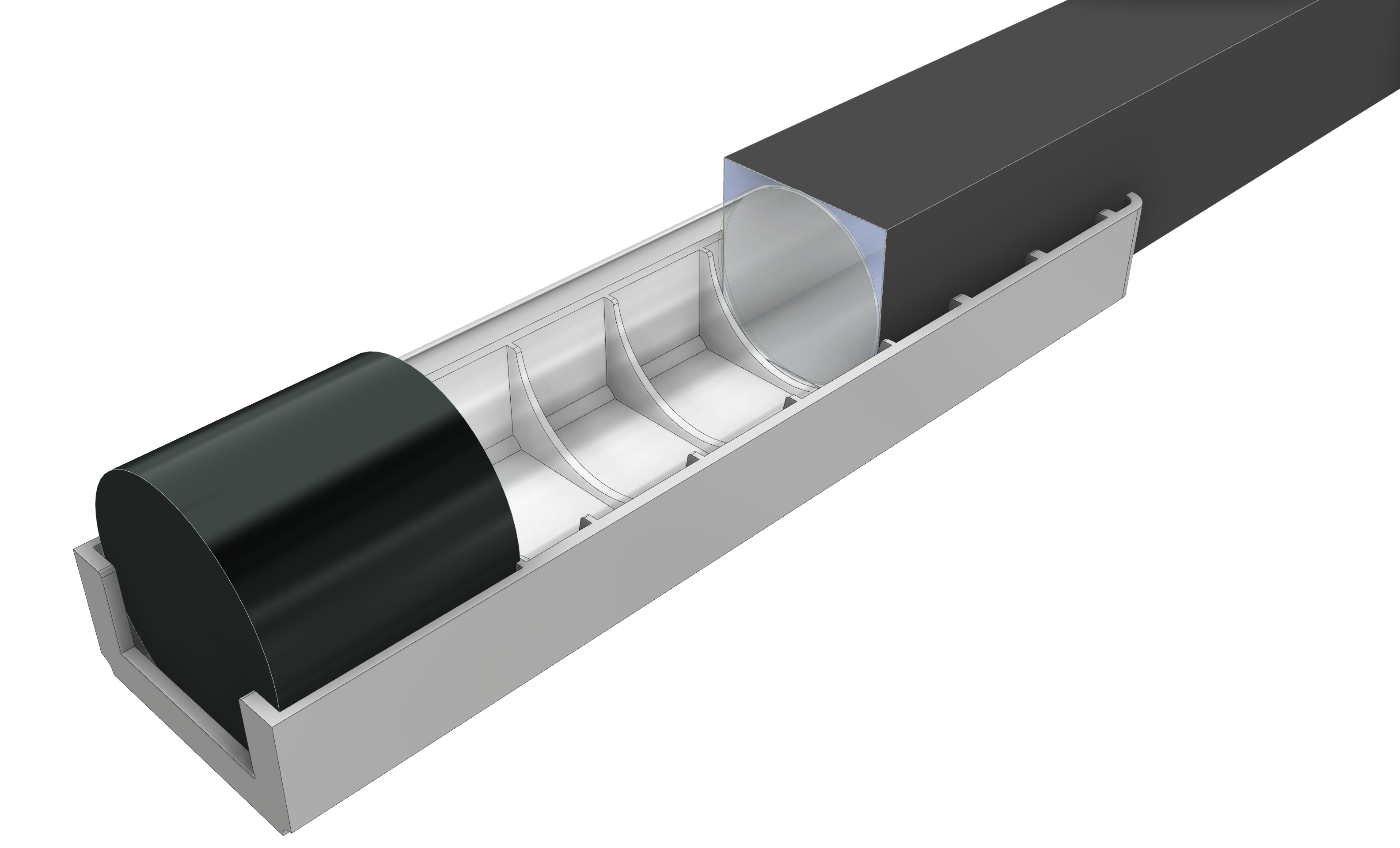} &
  \includegraphics[width=0.4\textwidth]{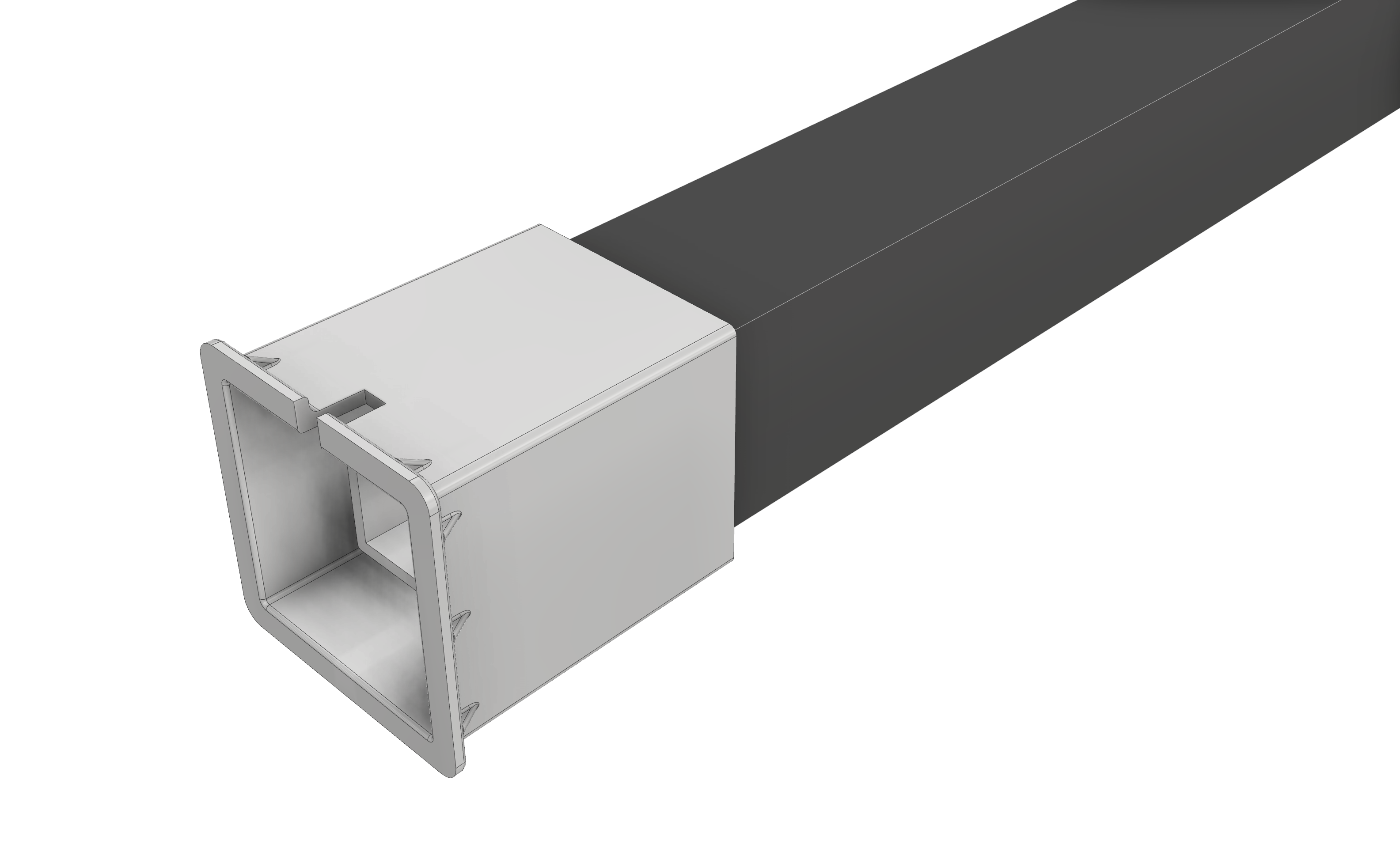} \\
  (a) & (b)
\end{tabular}
\caption{\label{fig:assem1} 3D models illustrating (a) the assembly of a \ac{pmt}, \ac{pmt} support, and scintillator, and (b) the attachment of the \ac{led} support to the opposite end of the scintillator. The \ac{pmt} is shown unwrapped for clarity.}
\end{figure}
\begin{figure}[h]
\centering
\includegraphics[width=0.39\textwidth]{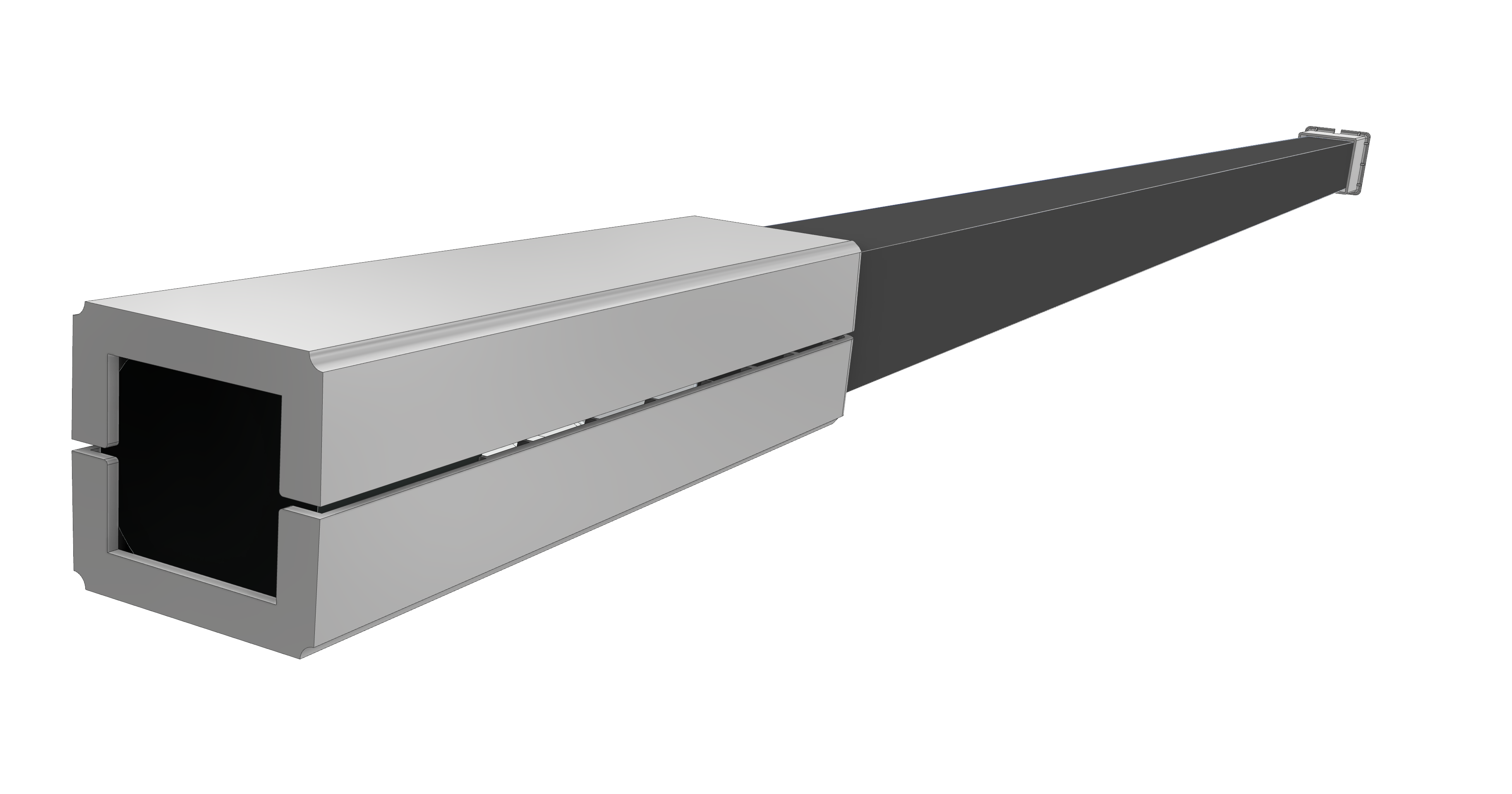}
\includegraphics[width=0.6\textwidth]{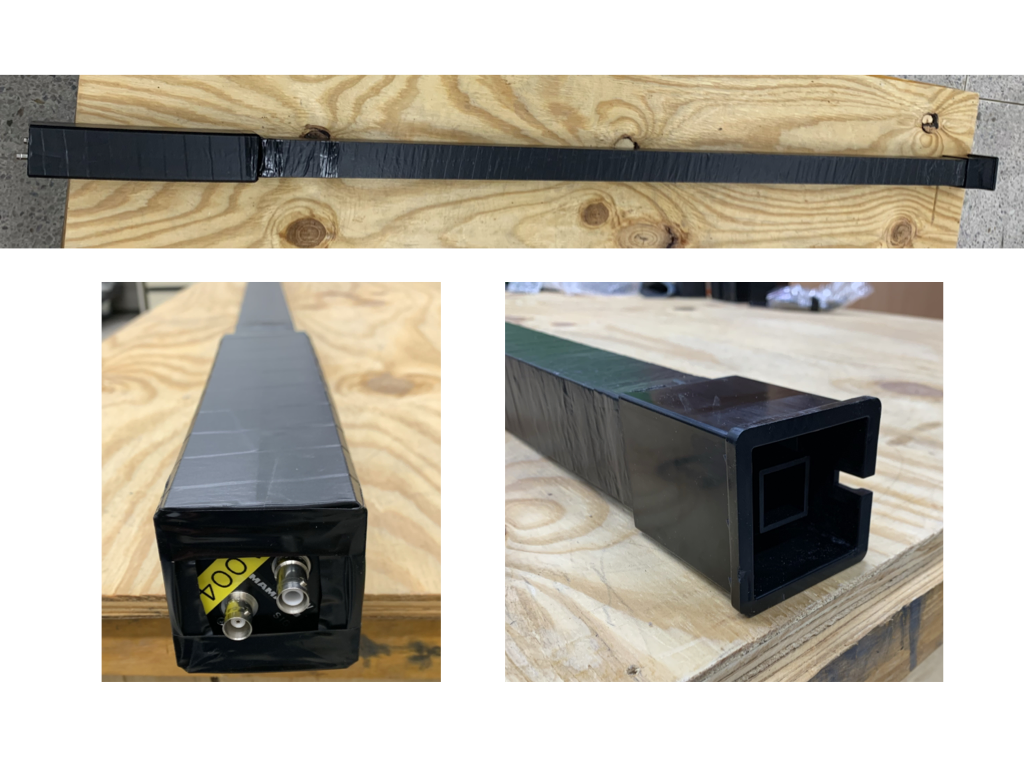}
\caption{\label{fig:assem2} A 3D model (left) and corresponding photographs (right) of a completed module.}
\end{figure}

Before assembling a module, the scintillator and \ac{pmt} are pre-wrapped. The \ac{pmt} is attached to the scintillator using optical grease to enhance the optical coupling between them. The \ac{pmt} support is then placed over the entire \ac{pmt} and approximately 100~mm of the scintillator to ensure secure mechanical coupling. Figure~\ref{fig:assem1}(a) shows the arrangement of the \ac{pmt} support, the \ac{pmt}, and the scintillator. Figure~\ref{fig:assem1}(b) shows the assembled \ac{led} support. We define a single module as the combination of the wrapped \ac{pmt} and the scintillator with their respective supports. A fully assembled module is shown in Figure~\ref{fig:assem2}.

\subsection{Magnetic shielding}\label{sec:mag_shielding}
In the presence of a magnetic field, the gain of a \ac{pmt} can decrease due to the deflection of photoelectrons emitted from the photocathode, which may prevent them from reaching the first dynode. This effect is most pronounced when the magnetic field is oriented perpendicular to the \ac{pmt} axis. Therefore, shielding \ac{pmt}s from external magnetic fields is essential to maintain high detection efficiency for single-photon signals.

\begin{figure}[htp]
\centering
\includegraphics[width=0.5\textwidth]{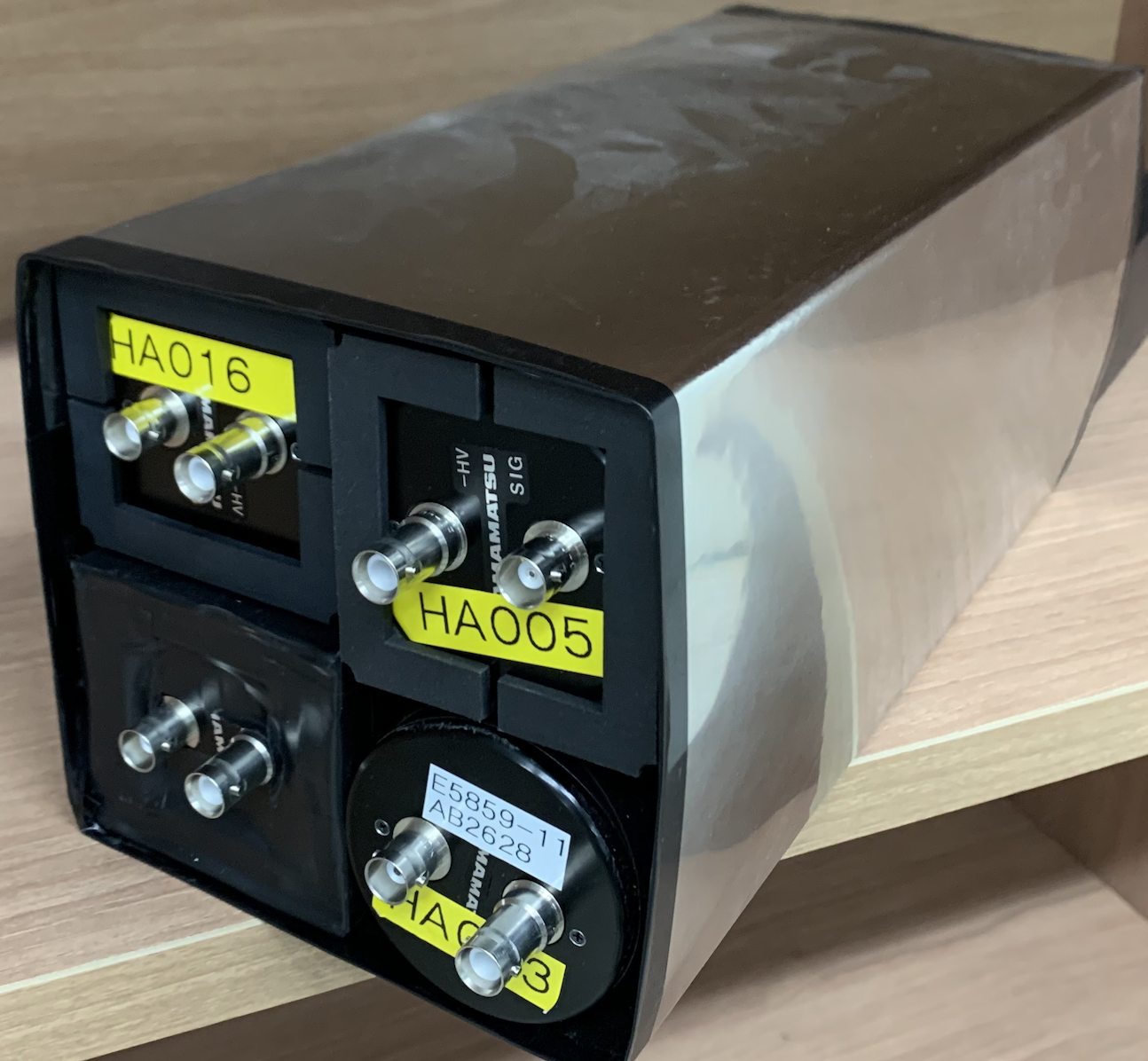}
\caption{\label{fig:mumetal} Four modules en-closed by mu-metal shielding (silver plates).}
\end{figure}
To shield against magnetic fields, the four \ac{pmt}s in each supermodule are enclosed in two layers of 0.2-mm-thick mu-metal plates, as shown in Figure~\ref{fig:mumetal}. The external magnetic field perpendicular to the \ac{pmt} axis is attenuated by more than 95\% inside the shielding. The impact of this reduction on \ac{pmt} gain was evaluated by comparing the pulse heights generated by cosmic muons. It was confirmed that the pulse height remains unaffected by magnetic fields of up to 2.5~mT, which is much larger than the magnetic field event when nearby detectors are in operation.

\section{Mechanical Support}\label{sec:mech_sup}
In order to control backgrounds while enhancing sensitivity to millicharged particles, the 80~modules are arranged in a \(10 \times 8\) configuration in each of the two layers. The axis along the long side of the modules is tilted by \(4.5^\circ\) with respect to the floor to ensure alignment with the target. The mechanical support structure is designed to meet these requirements, facilitate simple installation, and provide access to key module components for maintenance. The design consists of three main components: the ``Supermodule,'' the ``Cage,'' and the ``Table.''

\subsection{Supermodule}\label{sec:mech_supermodule}
A supermodule consists of four (\(2 \times 2\)) modules and a supermodule holder. The supermodule holder includes two brackets (\(130~\textrm{mm} \times 130~\textrm{mm} \times 20~\textrm{mm}\)) each featuring four square windows (\(55~\textrm{mm} \times 55~\textrm{mm}\)), three rods with diameters of 6~mm and lengths of 310~mm, 710~mm, and 750~mm, two disk-shaped end stoppers (diameter 50~mm), and a handle. All these components are made of aluminum. ISO M6 threaded holes are drilled at the center of the brackets, end stoppers, and handle, while ISO M6 threaded heads are present on both ends of the rods. All parts of the supermodule holder are connected through these ISO M6 threads.
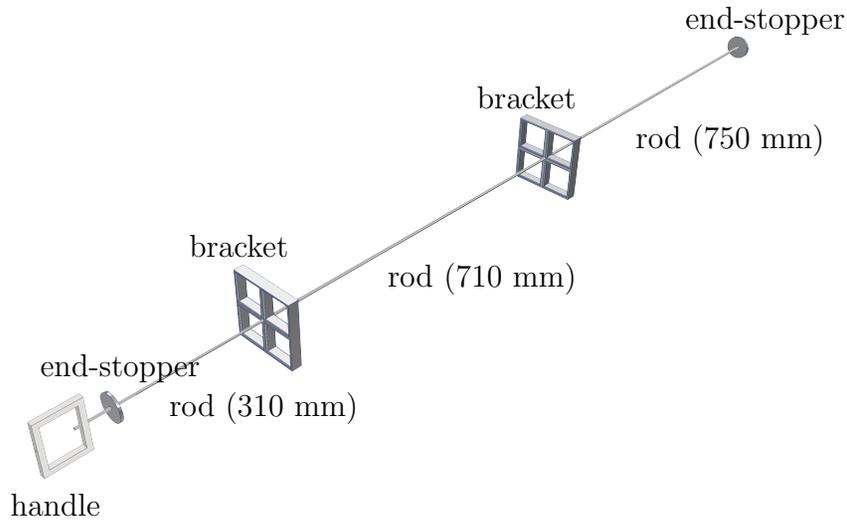
\begin{figure}[h]
\begin{center}
\begin{overpic}[width=0.8\textwidth]{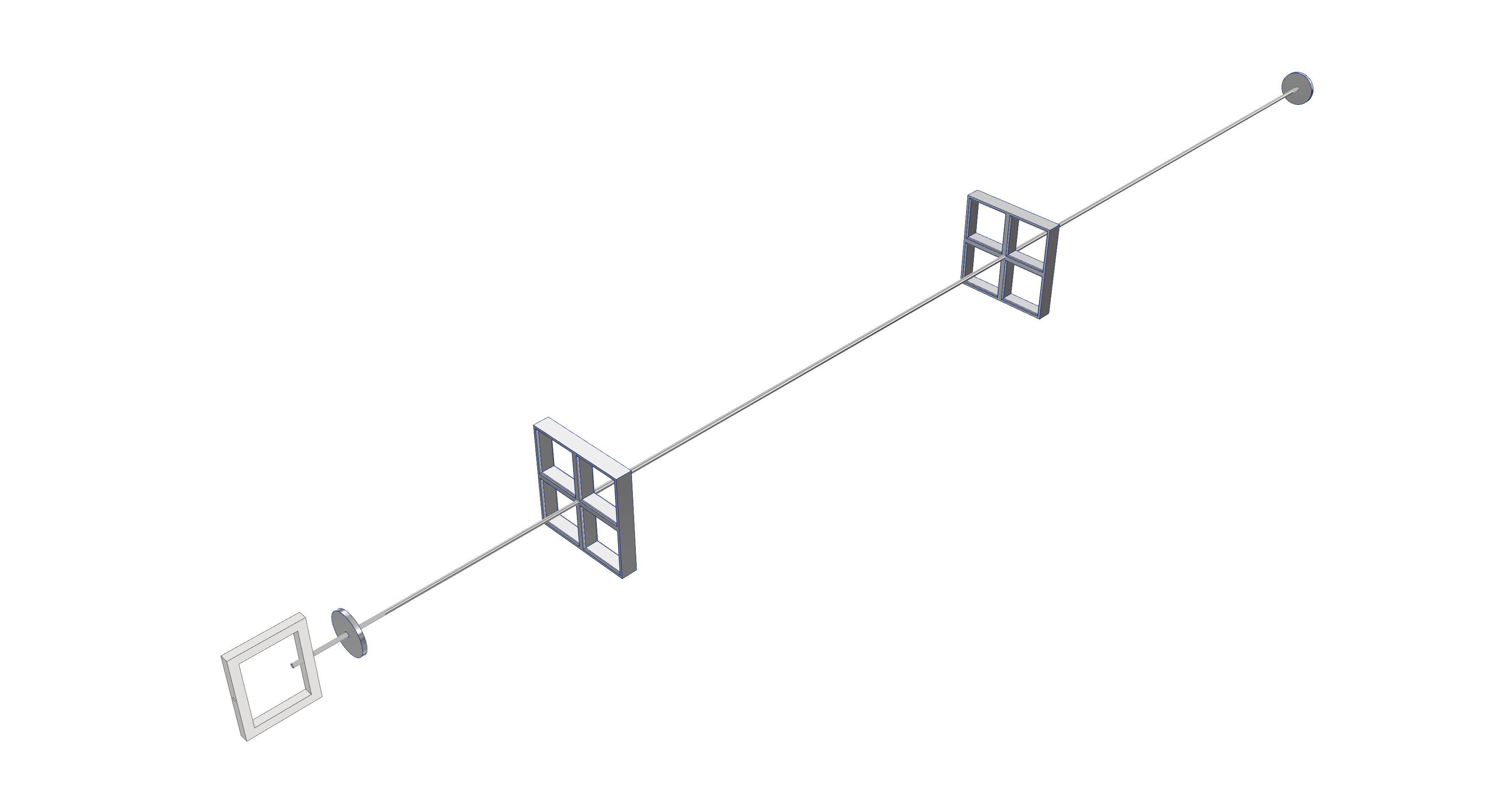}
  \put(13, 1){handle}
  \put(16,15){end-stopper}
  \put(29,11){rod (310~mm)}
  \put(31,27){bracket}
  \put(51,24){rod (710~mm)}
  \put(60,42){bracket}
  \put(76,38){rod (750~mm)}
  \put(81,50){end-stopper}
\end{overpic}
\end{center}
\caption{Isometric view of the supermodule holder structure.}
\label{fig:supermoduleholder}
\end{figure}
The 3D model of the supermodule holder is shown in Figure~\ref{fig:supermoduleholder}. The brackets hold four (\(2 \times 2\)) modules in place, while the two end-stoppers secure the modules to the supermodule holder and apply inward pressure to enhance the coupling between the \ac{pmt} and the scintillator.
\begin{figure}[h]
\begin{center}
\begin{overpic}[width=0.15\textwidth]{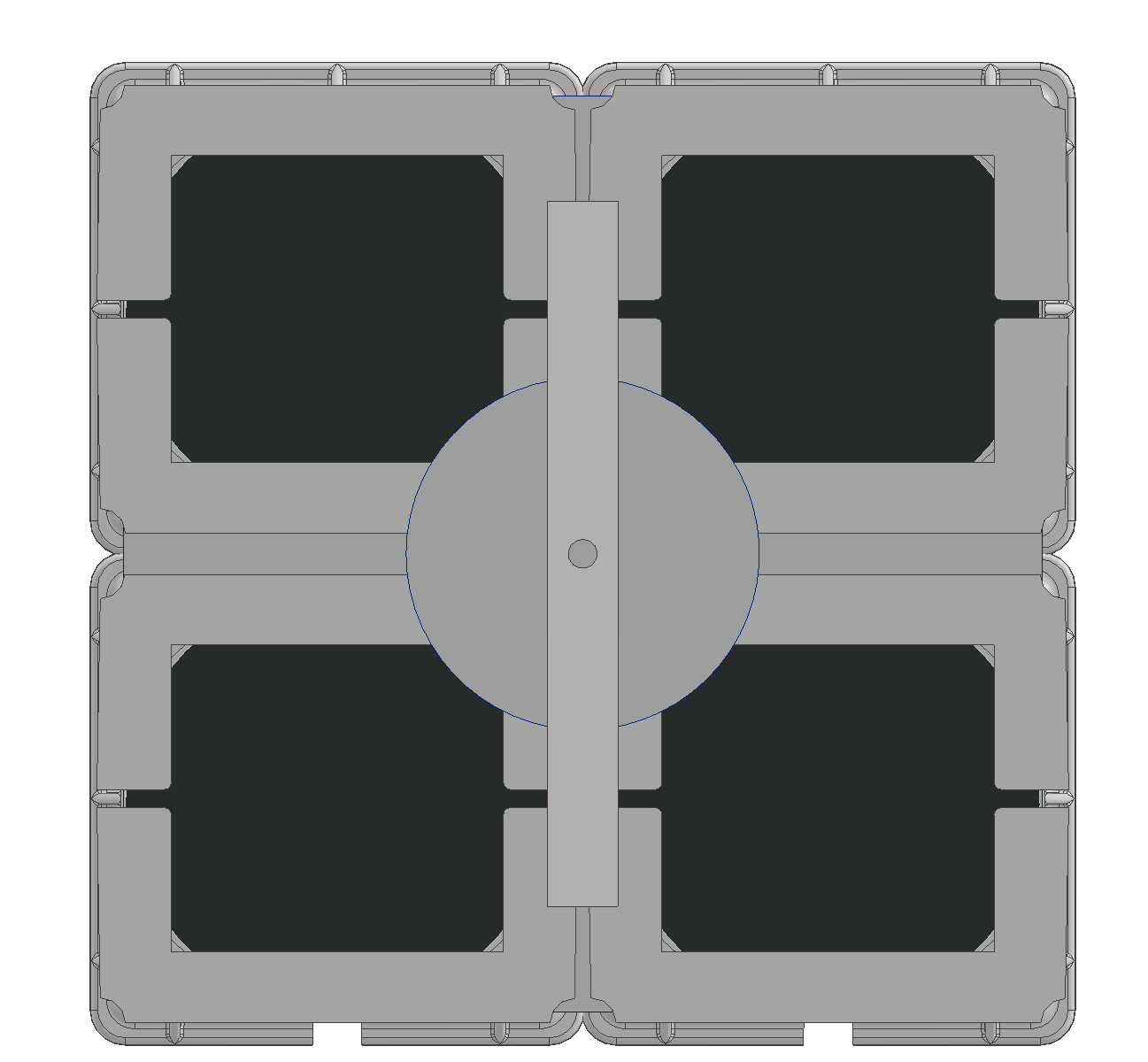}
  \put(20, 120){end-stopper }
  \put(20, 100){+ handle}
\end{overpic}
\begin{overpic}[width=0.15\textwidth]{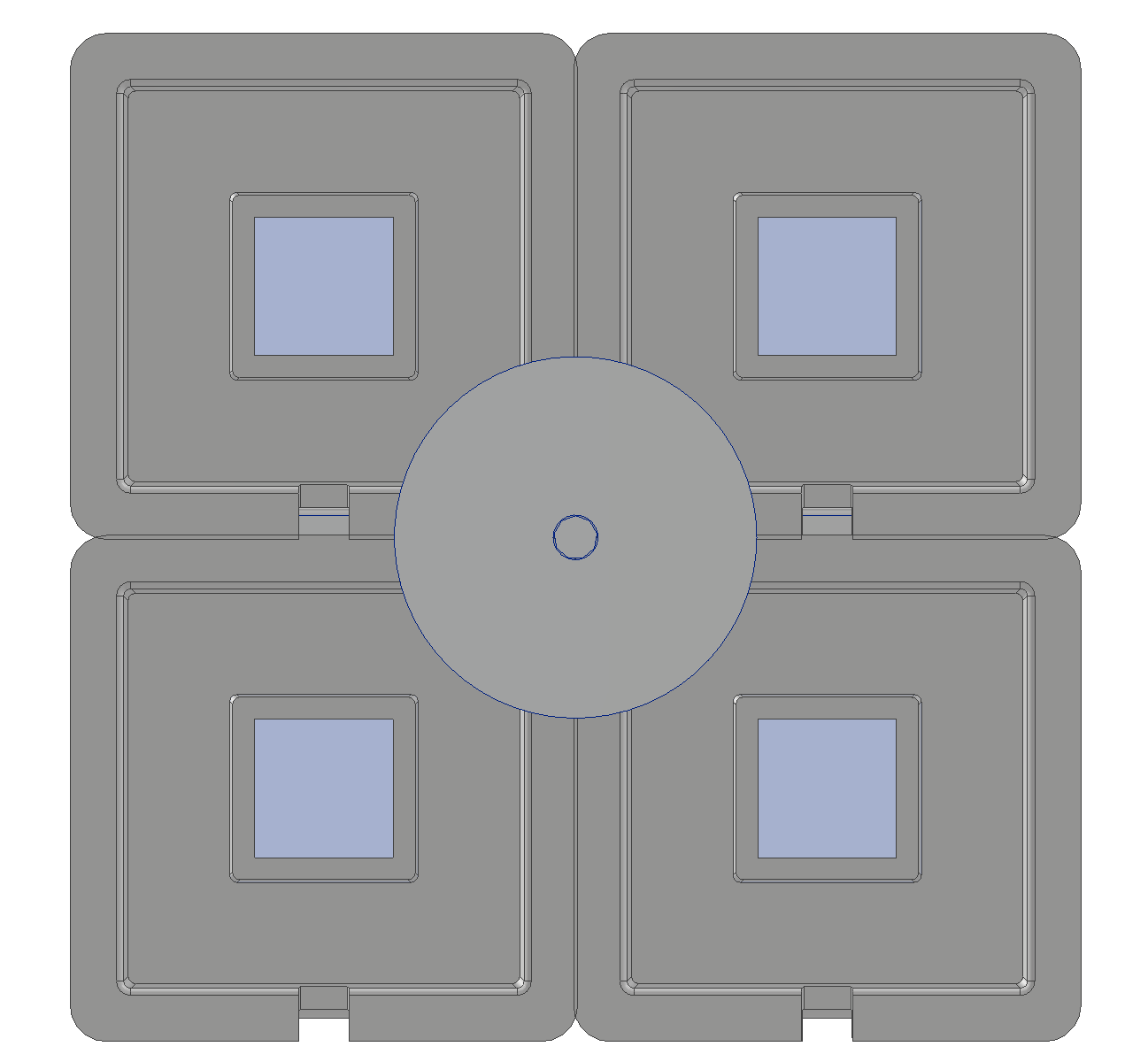}
  \put(27, 100){end-stopper}
\end{overpic}
\begin{overpic}[width=0.59\textwidth]{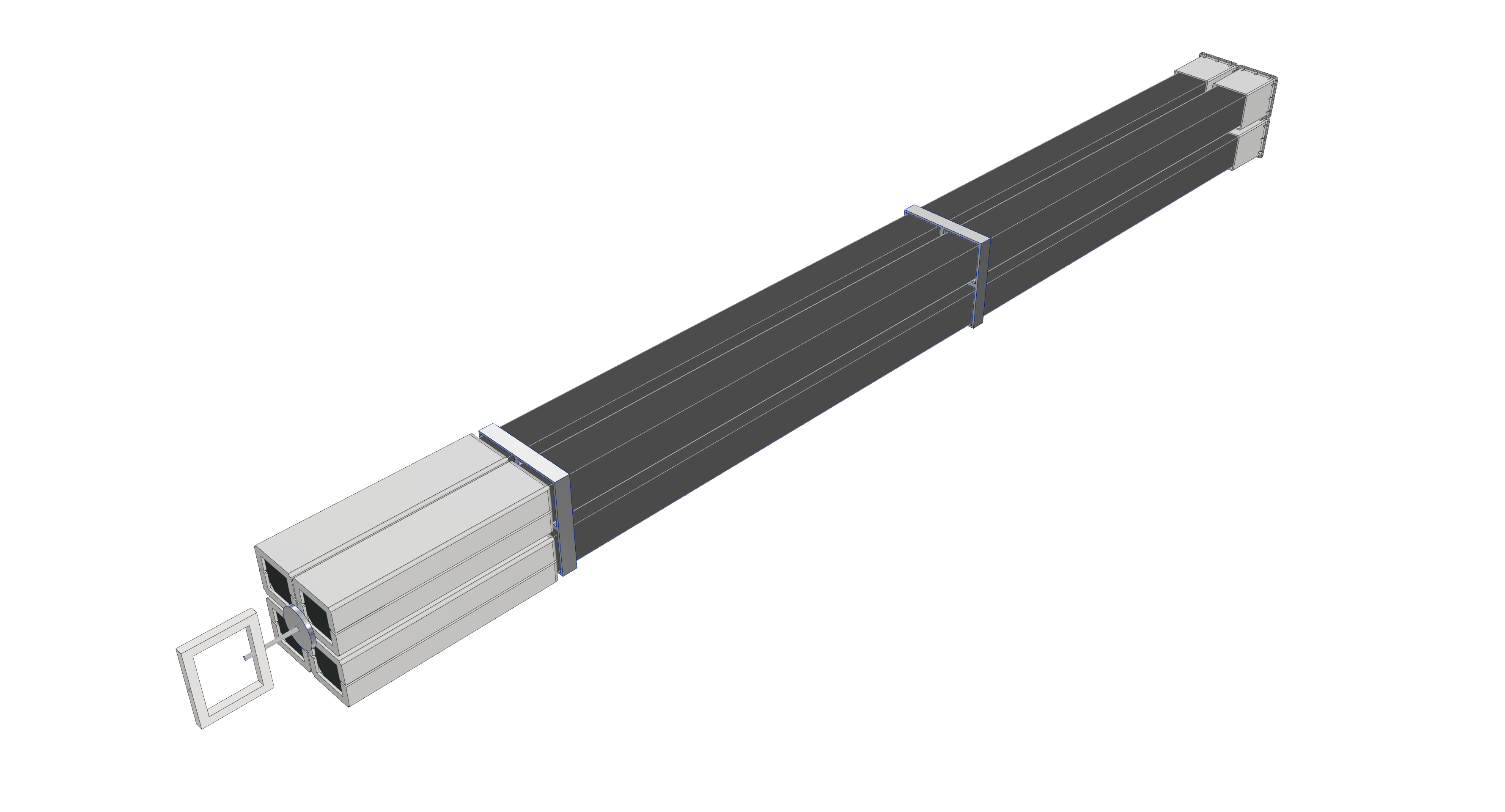}
  \put(11, 2){handle}
  \put(12, 20){\ac{pmt} support}
  \put(38, 13){bracket}
  \put(65, 29){bracket}
  \put(82, 40){\ac{led} support}
\end{overpic}
\end{center}
\caption{Views of a supermodule from the \ac{pmt} support side and the \ac{led} support side (left). Two end-stoppers secure the four ($2 \times 2$) modules within the supermodule holder. Isometric view illustrating the structure of a supermodule (right).}
\label{fig:supermodule}
\end{figure}
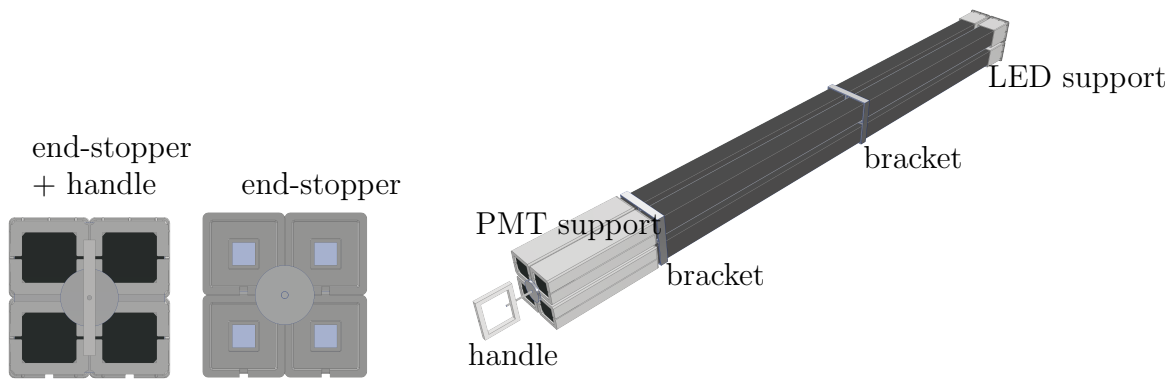
Side and isometric views of a supermodule are shown in Figure~\ref{fig:supermodule}, respectively.
\begin{figure}[htbp]
\centering
\includegraphics[width=0.8\linewidth]{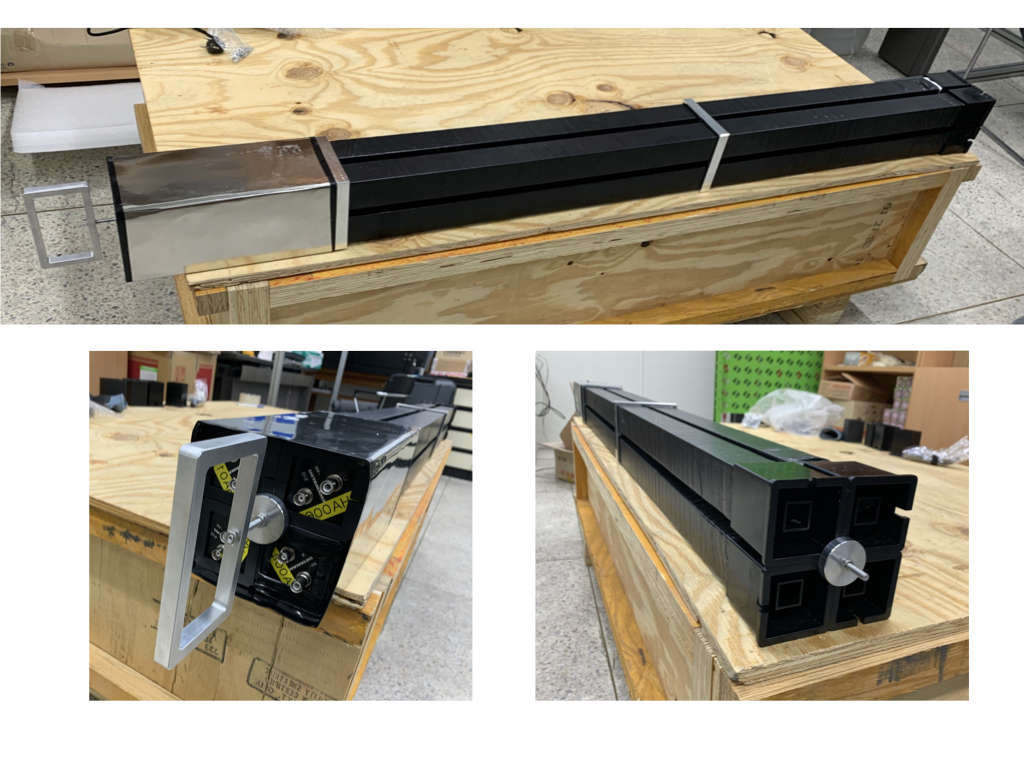}
\caption{Photographs of a fully assembled supermodule.}
\label{fig:photo_supermodule}
\end{figure}
Figure~\ref{fig:photo_supermodule} shows photographs of a completed supermodule with \ac{pmt} supports covered by mu-metal shielding.

\subsection{Cage}\label{sec:mech_cage}
\begin{figure}
\begin{center}
\begin{overpic}[width=0.25\textwidth]{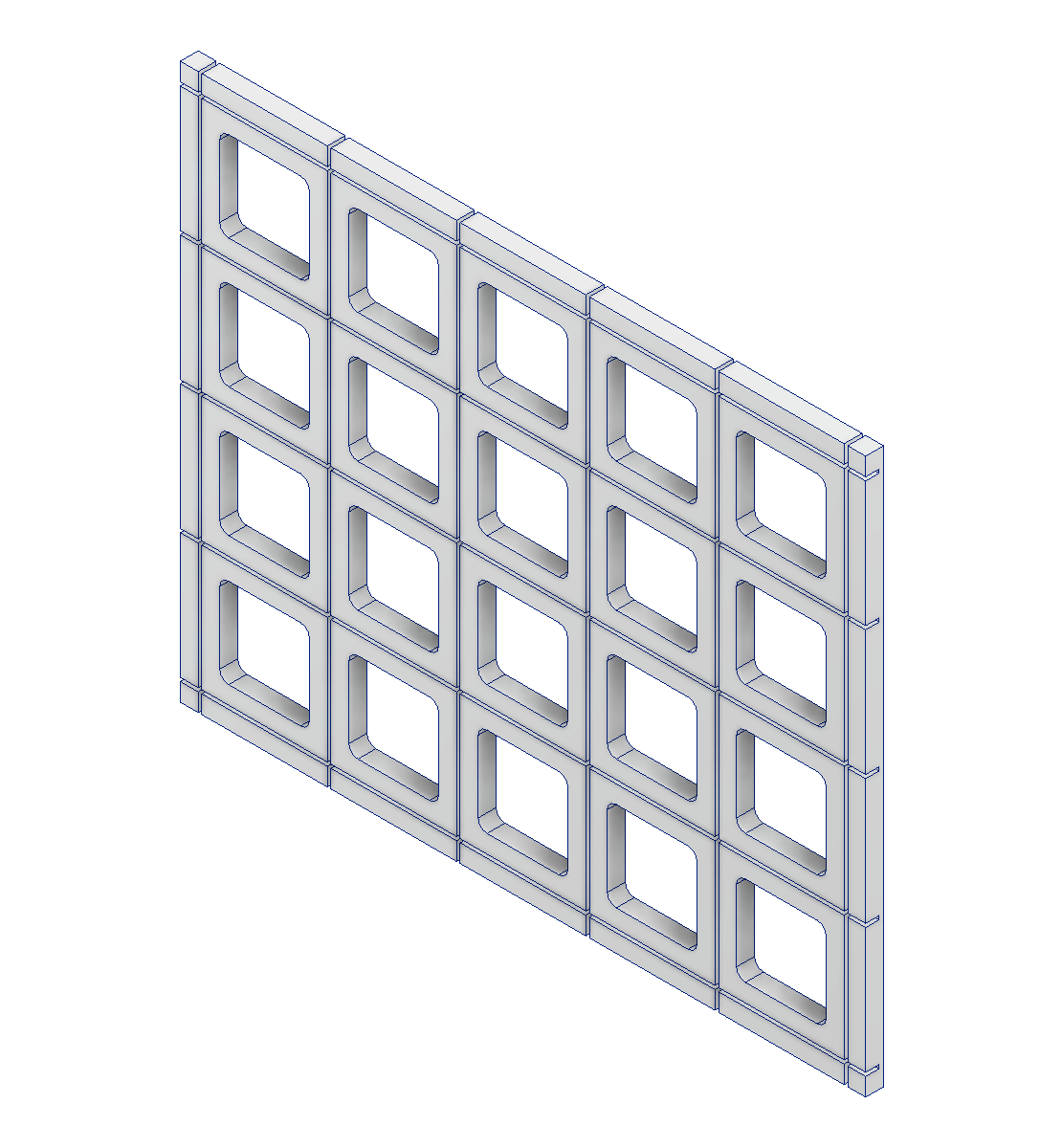}
\end{overpic}
\begin{overpic}[width=0.08\textwidth]{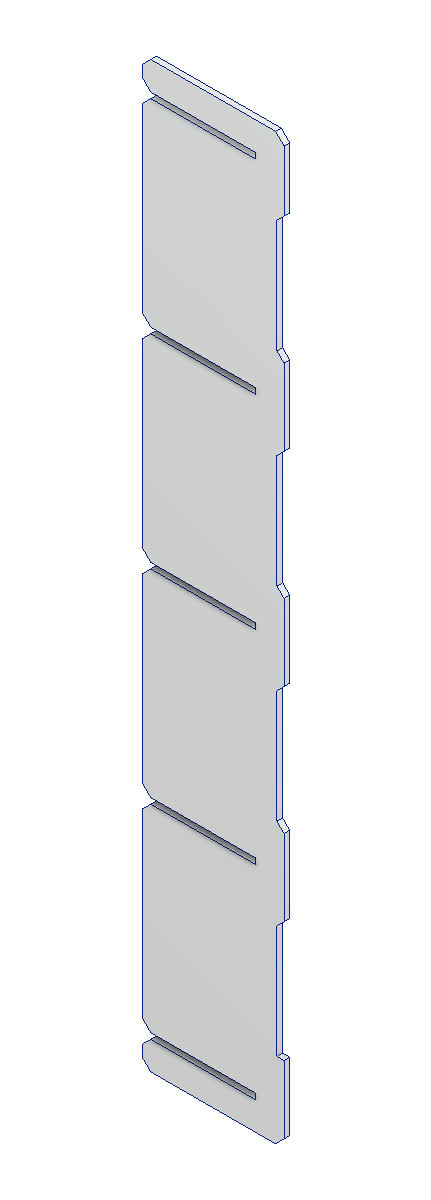}
\put(-8,70){slot}
\end{overpic}
\begin{overpic}[width=0.4\textwidth]{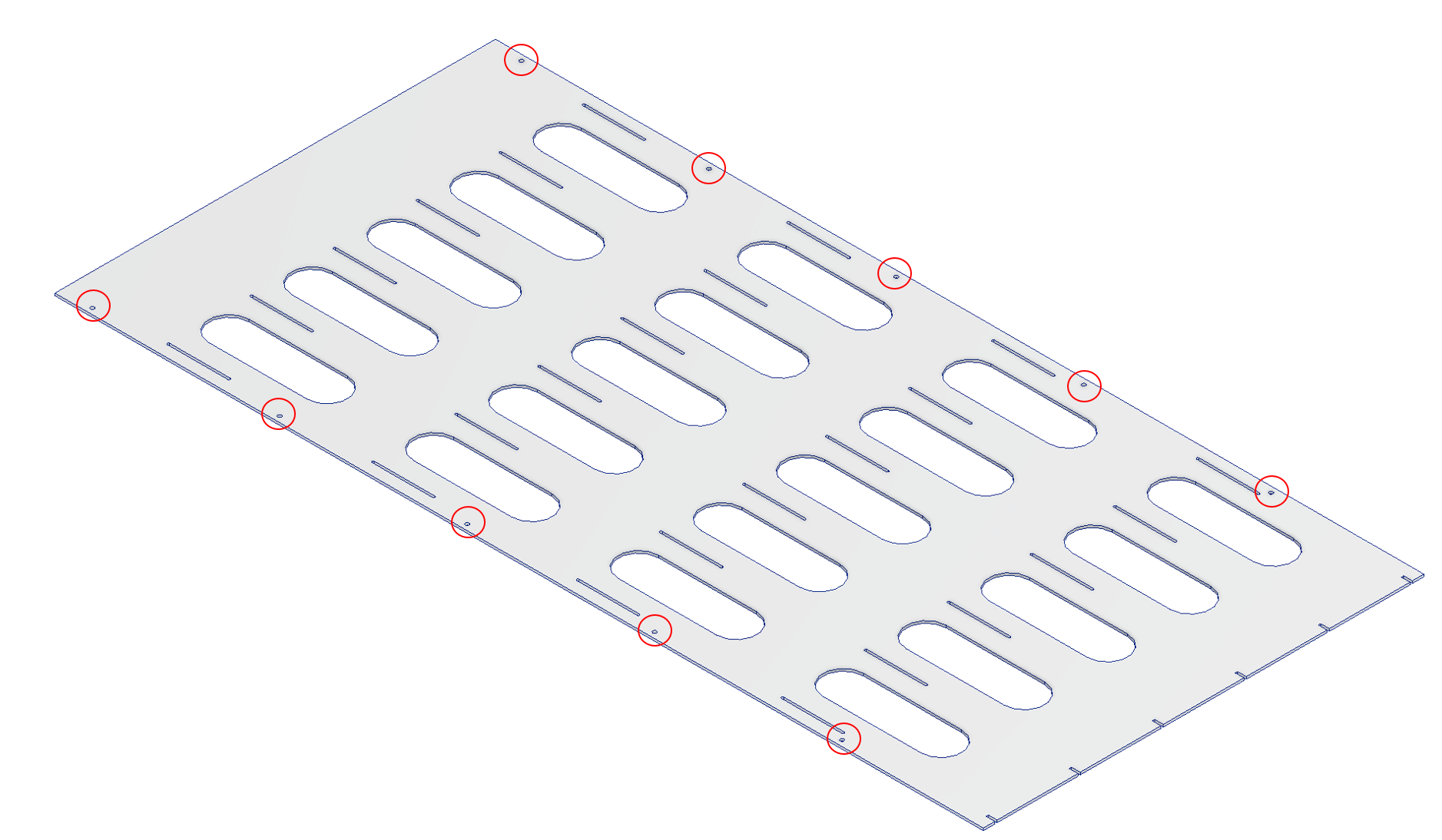}
\end{overpic}
\begin{overpic}[width=0.1\textwidth]{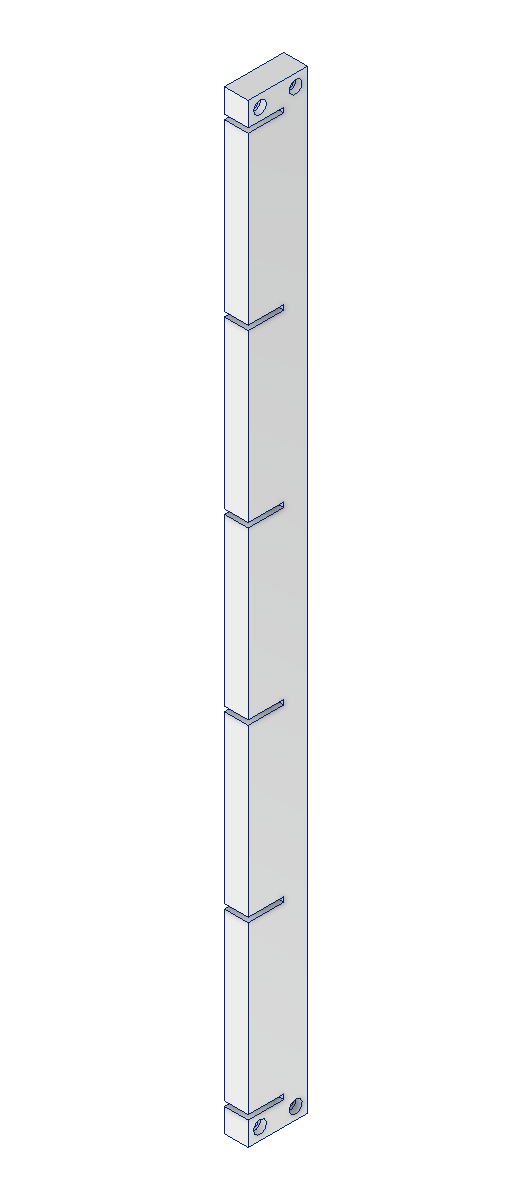}
\end{overpic}
\end{center}
\caption{From the left, isometric views of the end-cap, vertical plate, horizontal plate, and cage-cap.}
\label{fig:cage}
\end{figure}

The second component of the design is the Cage, which holds twenty supermodules (80 modules arranged in a \(5 \times 4\) configuration). It consists of four main aluminum parts: vertical plates (30 pieces), horizontal plates (5 pieces), cage-caps (8 pieces), and an end-cap (1 piece). As shown in Figure~\ref{fig:cage}, each vertical plate measures \(100~\mathrm{mm} \times 620~\mathrm{mm} \times 4~\mathrm{mm}\) and has five slots, each 4 mm wide and 80 mm long, where the horizontal plates are connected. Each horizontal plate features \(5 \times 4\) oval holes to reduce weight and eight slots, each 105 mm long and 4 mm wide, for connection to the vertical plates. Ten ISO M6 threaded holes (indicated by red circles in Figure~\ref{fig:cage}) on each horizontal plate are used to secure the Cage to the Table. The cage cap is an aluminum bar measuring \(764~\mathrm{mm} \times 50~\mathrm{mm} \times 20~\mathrm{mm}\) that holds the vertical and horizontal plates (top and bottom) together. Four cage caps are used on both the top and bottom of the Cage.

After inserting all 20 supermodules into the Cage, four supermodules in each column are secured to the end cap using an aluminum bar called the ``secure bar.'' This bar has four holes designed for the aluminum rods from the supermodules to be screwed into.
\begin{figure}[h]
\begin{center}
\includegraphics[width=0.12\linewidth]{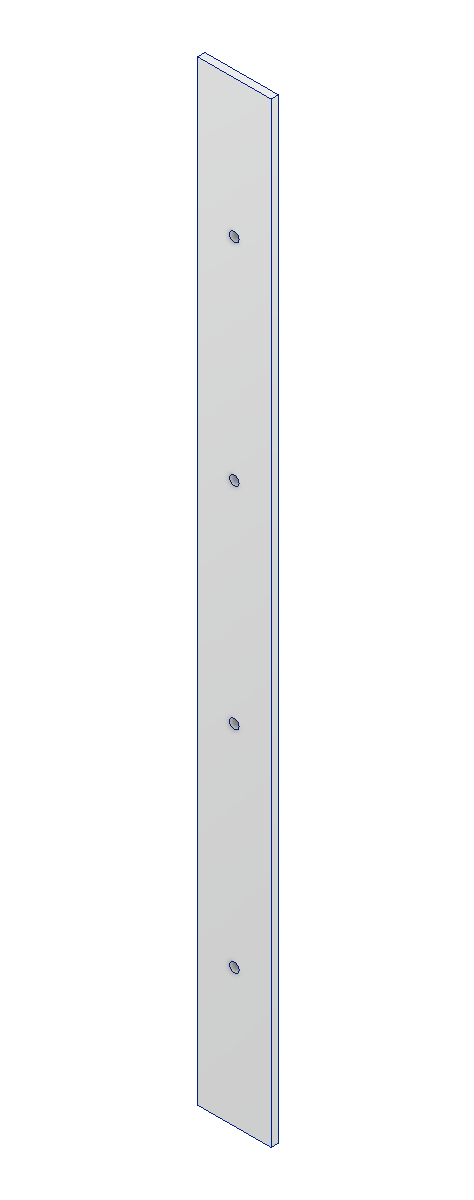}
\includegraphics[width=0.45\linewidth]{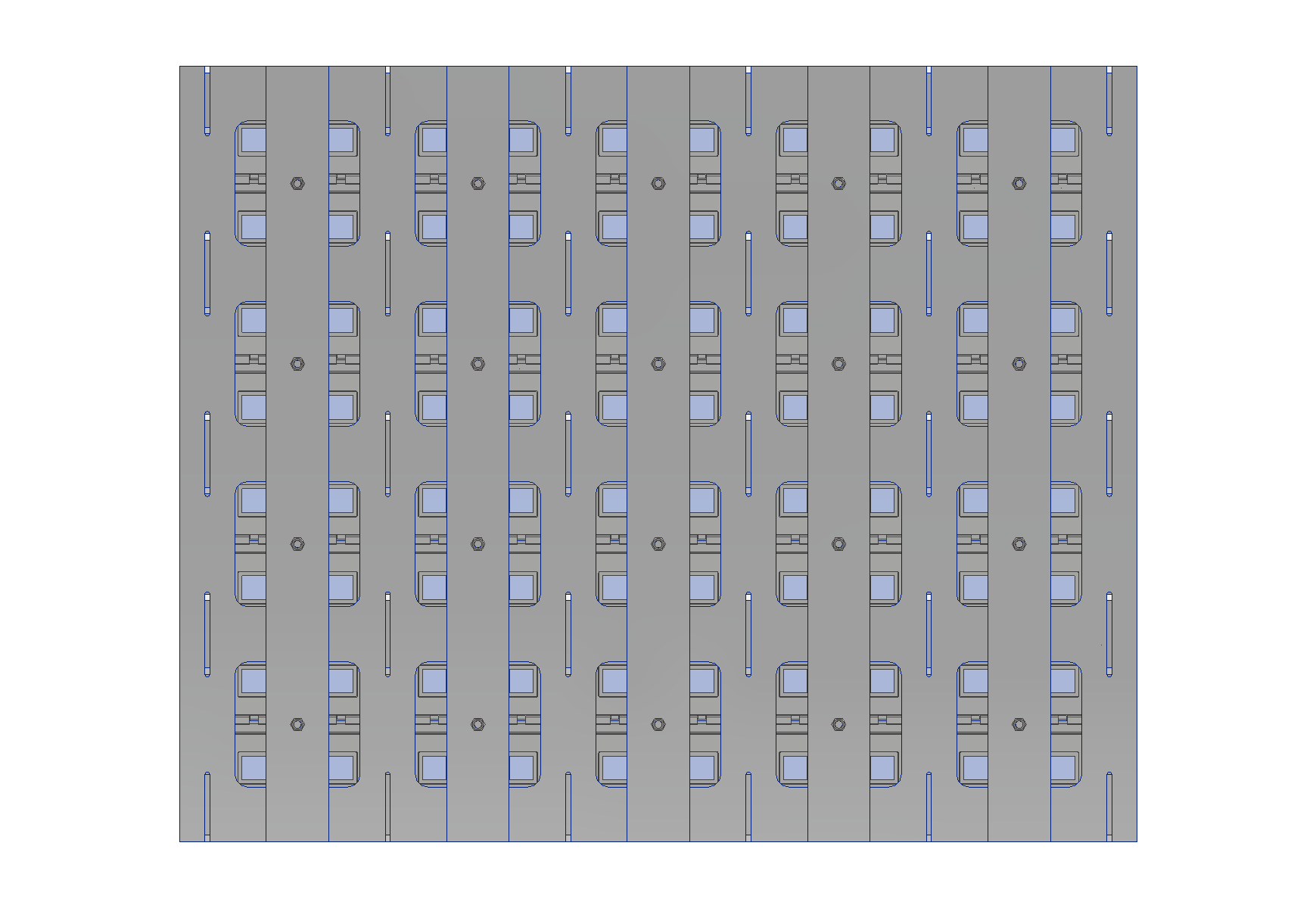}
\end{center}
\caption{Isometric view of a secure bar (left) and the Cage as seen from the \ac{led} support side (right). Five secure bars and the rods of the supermodule holders are connected via ISO M6 threads.}
\label{fig:securebar}
\end{figure}
Figure~\ref{fig:securebar} shows the secure bar and end-cap after the supermodules have been inserted into the Cage. The Cage, with supermodules installed, measures \(1850~\textrm{mm} \times 764~\textrm{mm} \times 620~\textrm{mm}\).
\begin{figure}[h]
\begin{center}
\includegraphics[width=0.75\linewidth]{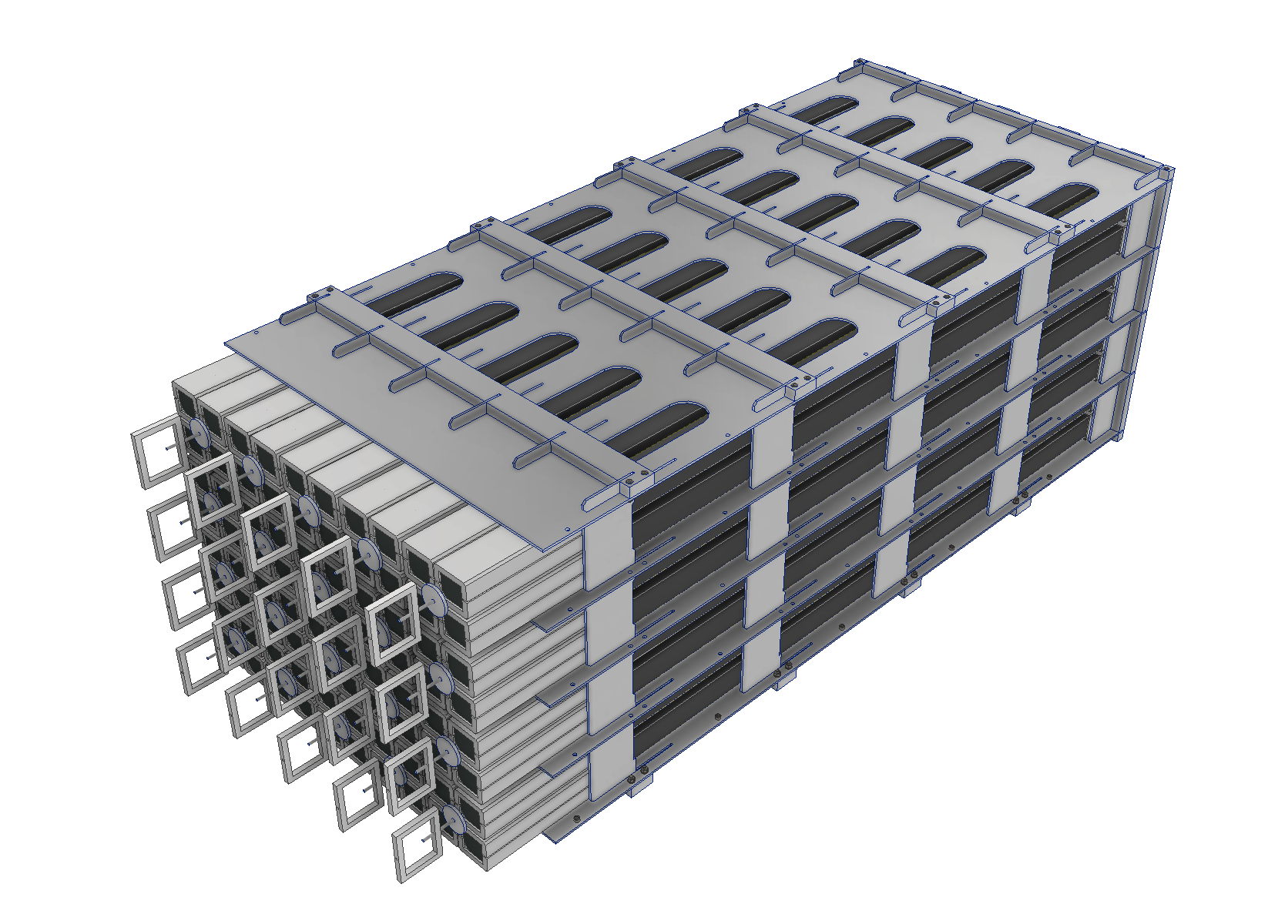}
\end{center}
\caption{Isometric view of the Cage fully assembled with $5 \times 4$ supermodules installed.}
\label{fig:cage_supermodules}
\end{figure}
An isometric view of the fully assembled Cage and supermodules is shown in Figure~\ref{fig:cage_supermodules}.

\subsection{Table}\label{sec:mech_table}
The final component of the design is the ``Table.'' There are two Tables in the \ac{submet} detector: ``Table 1'' (the near side to the target) and ``Table 2'' (the far side from the target). The Tables are constructed using 4040 (\(40~\textrm{mm} \times 40~\textrm{mm}\)) and 5050 (\(50~\textrm{mm} \times 50~\textrm{mm}\)) aluminum profiles, along with L-shaped aluminum brackets.

\begin{figure}[h]
\begin{center}
\begin{overpic}[width=0.5\textwidth]{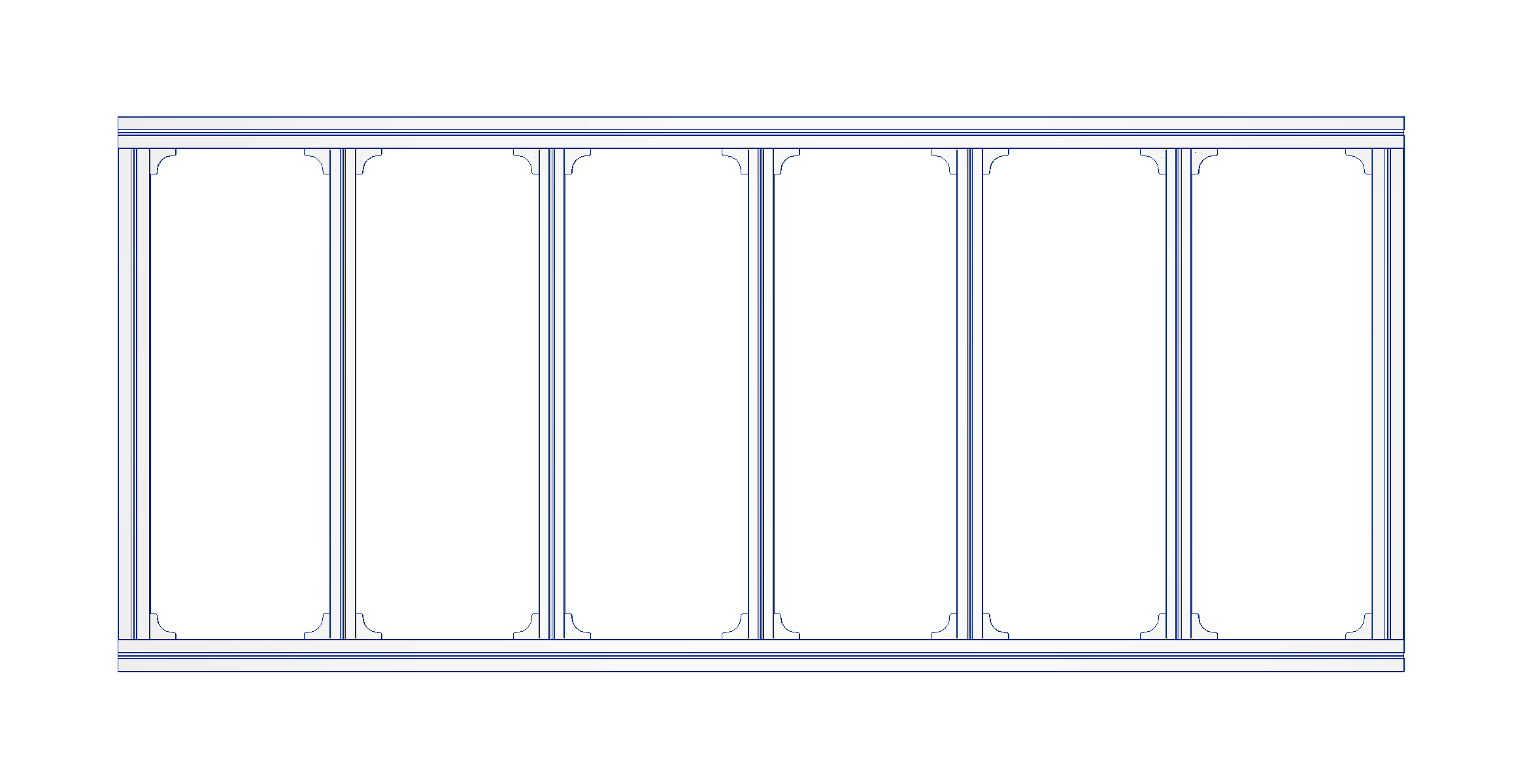}
  \put(35,45){5050 profile}
  \put(93,25){5050 profile}
  \put(35,25){4040 profiles}
\end{overpic}
\begin{overpic}[width=1.0\textwidth]{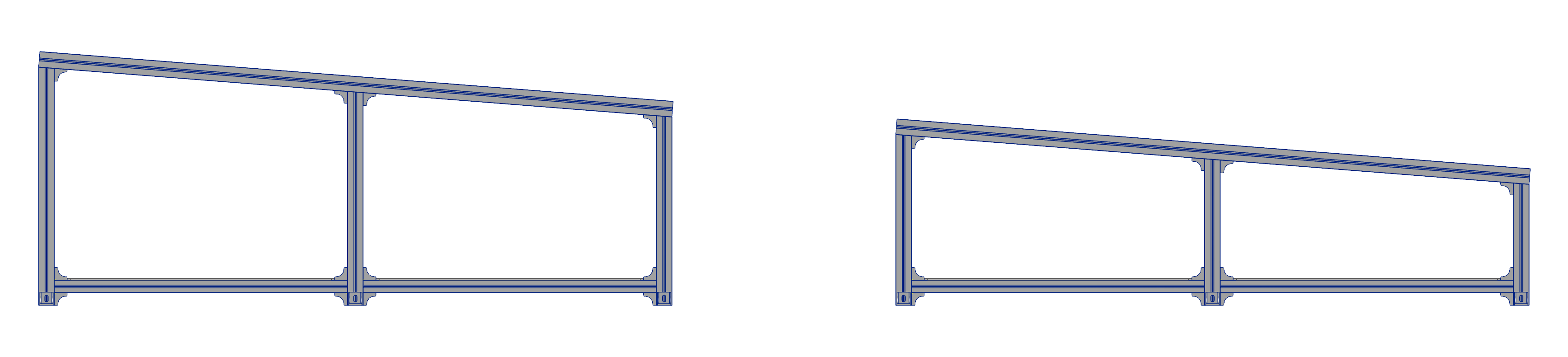}
  \put(22,20){Table 1}
  \put(71,16){Table 2}
  \put(0,0){750~mm}
  \put(92,0){381~mm}
\end{overpic}
\end{center}
\caption{Top view of the Table (upper panel) and side views of Table 1 and Table 2 (lower panel).}
\label{fig:table}
\end{figure}
Figure~\ref{fig:table} illustrates the rectangular frame formed by the 5050 profiles, with the 4040 profiles placed inside. The dimensions are chosen so that a Cage fits precisely within the rectangle created by the 5050 profiles.

The $4.5^\circ$ angle between the Cages and the floor is achieved by using legs of different lengths for the Tables. The height of the legs on the near side (towards the target) is 750~mm, while those on the opposite end measure 381~mm, as shown in Figure~\ref{fig:table}. ISO M6 threads are used to secure the Cage to the Table. A Cage placed on a Table is referred to as a ``Layer.'' The layers on the near and far sides relative to the target are designated as Layer 1 and Layer 2, respectively.

\begin{figure}[h]
\begin{center}
\includegraphics[width=0.65\linewidth]{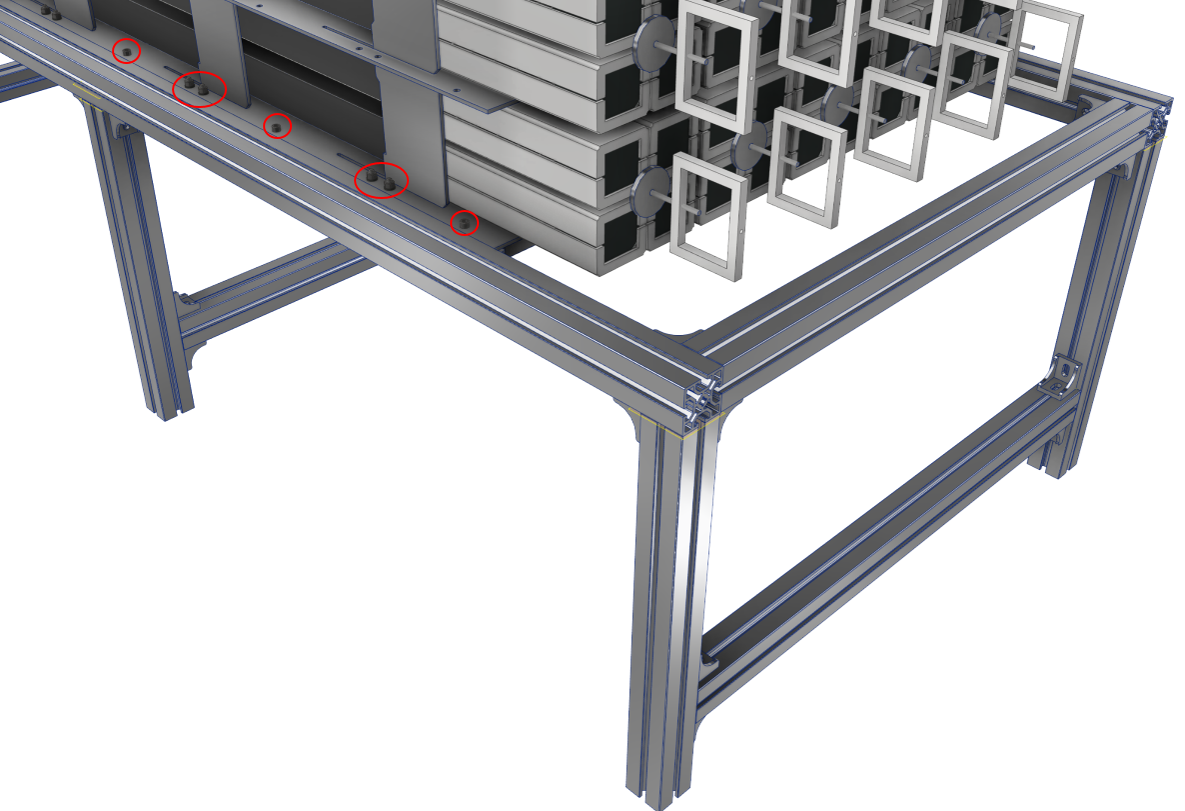}
\end{center}
\caption{Isometric view of a Layer. Red circles indicate the locations of ISO M6 threaded holes used to secure the bottom horizontal plate of the Cage to the 4040 profiles of the Table.}
\label{fig:table_fix}
\end{figure}

\subsection{Detector Installation}\label{sec:detector_installed}
\begin{figure}[h]
\begin{center}
\includegraphics[width=0.99\linewidth]{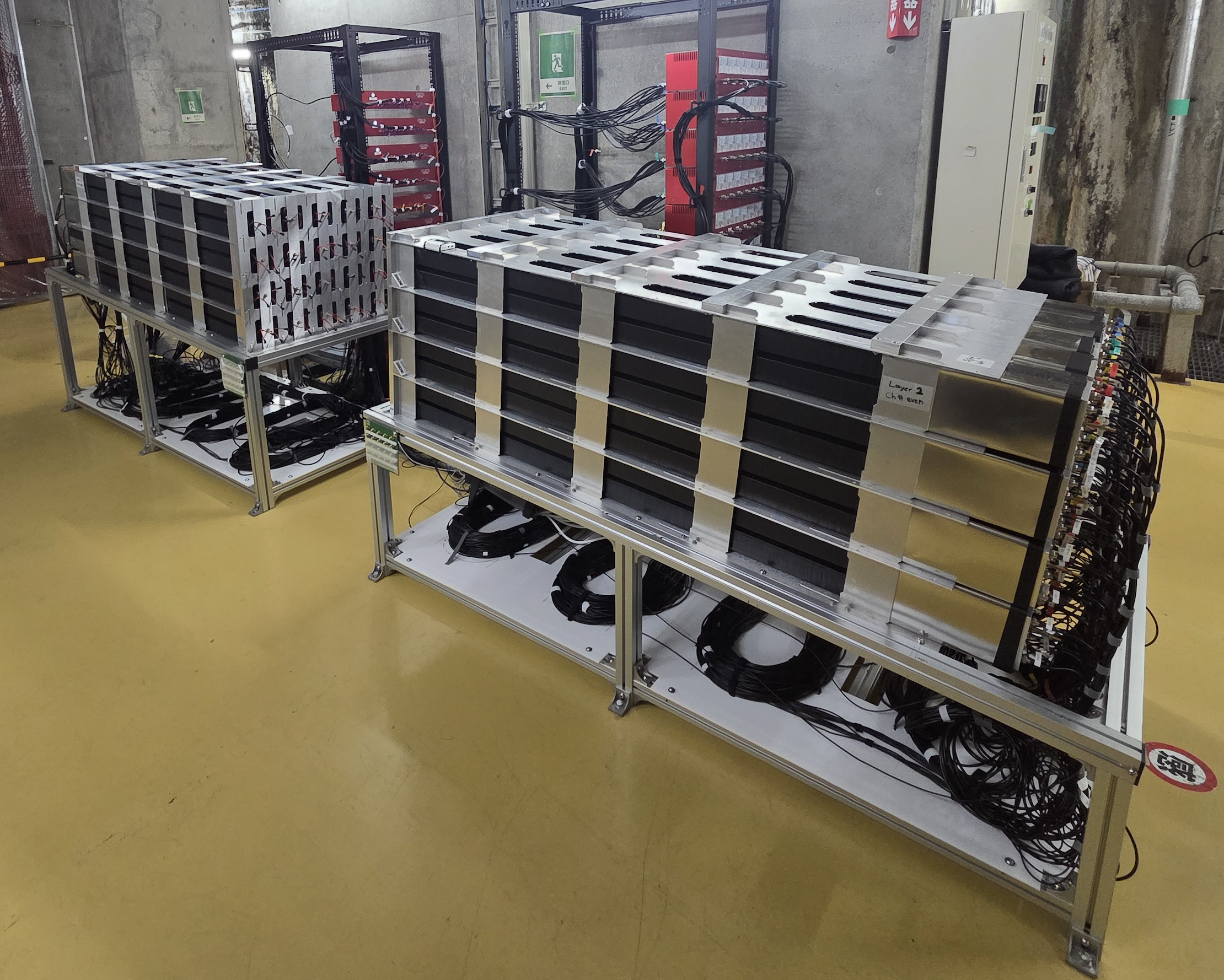}
\end{center}
\caption{Side view of the installed detector.}
\label{fig:detector_installed}
\end{figure}
Each of the detector components, as well as the cages and tables, was built as described in the previous sections at Korea University. The detector was then fully assembled to validate the complete system. The functionality of all 160 modules was confirmed by successfully identifying and reconstructing $O(100)$ cosmic muon tracks. Subsequently, the detector was disassembled, transported, and reassembled at the detector site in May 2024. Figure~\ref{fig:detector_installed} shows a side view of the detector installed on the B2 floor of the Neutrino Monitor building.

\section{Weight and seismic analyses}\label{sec:fea}
The structural stability of the \ac{submet} detector, both under its own weight and during seismic events, is critical for the safe and reliable operation of the experiment. To assess this, we conducted static (weight) and dynamic (seismic) analyses to evaluate the structural integrity of the detector under these conditions.

\subsection{Weight Analysis}\label{sec:weight}
A static analysis was conducted to evaluate the stress exerted on the mechanical structure, the corresponding safety factor, and the maximum displacement of the table segments under static load. The analysis utilized the \textit{Autodesk Inventor Professional 2025} software~\cite{inventor}, with 3D CAD models of the Tables serving as input. The total weight of the mechanical supports (Table 1 and Table 2) was considered as the dead load\footnote{A constant load in a structure due to the weight of the members, the supported structure, and permanent attachments or accessories. In this context, the dead load refers to the weight of the Tables themselves.}, and a live load\footnote{A live load refers to a load that can change over time, such as movable objects. The table bears a variable load depending on the number of supermodules. Although modules exert a constant load after installation, we treat these as live loads to simplify the modeling of additional loads applied to the Table.} of 6.5~kN for each Table was applied in the simulation. The 6.5~kN live load accounts for the weight of the Cage, and supermodules.

\begin{figure}[h]
\begin{center}
\includegraphics[width=0.8\linewidth]{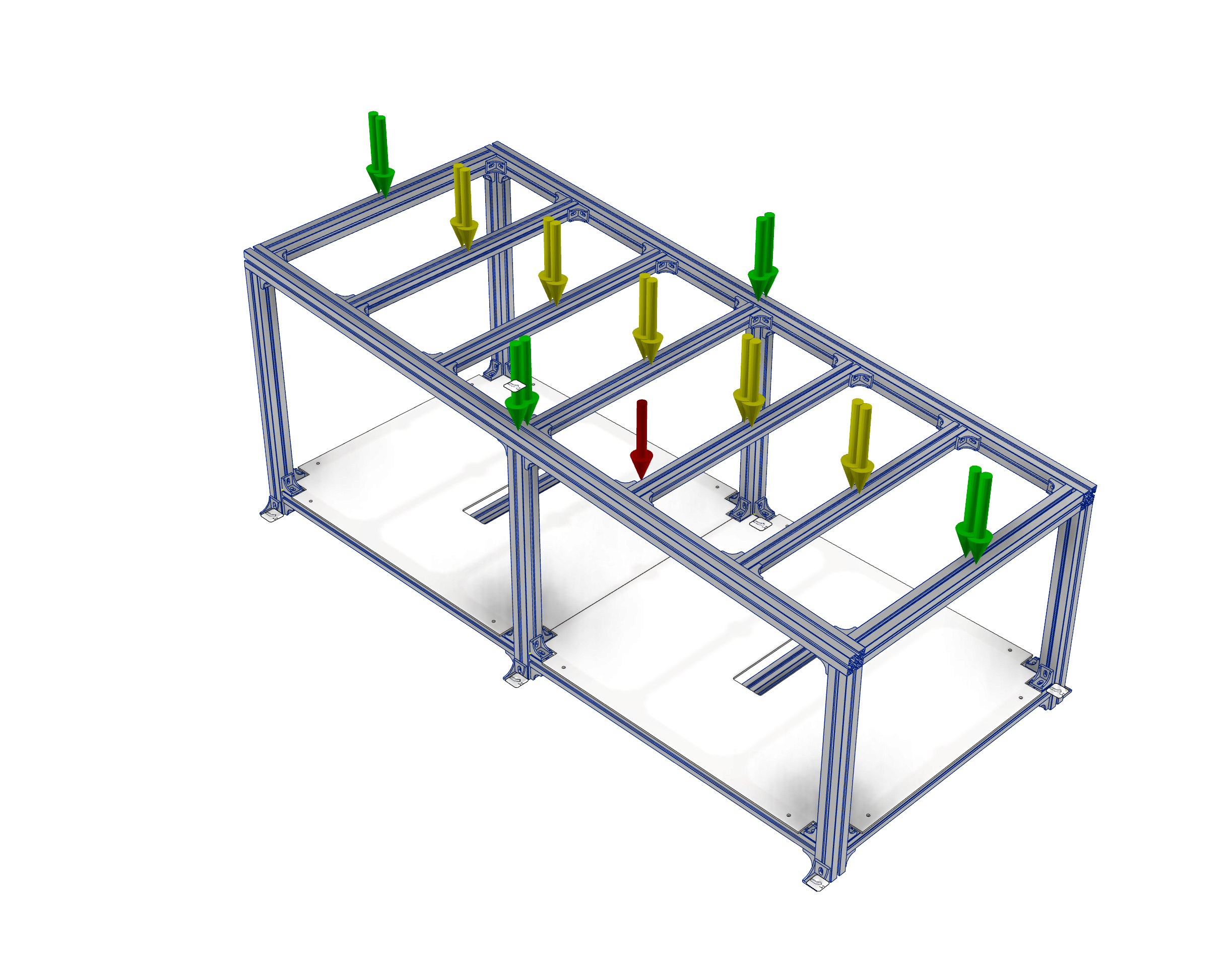}
\end{center}
\caption{Visualization of the applied loads for the weight analysis. Red arrows represent gravitational forces. Yellow arrows indicate uniform pressure applied to the central frames due to the Cage weight of 4.5 kN. Green arrows denote uniform pressure applied to the edge frames accounting for an additional 2 kN load from potential detector upgrade components.}
\label{fig:weight_load}
\end{figure}
The types and magnitudes of the loads applied to the structure are illustrated in Figure~\ref{fig:weight_load}. The total mass of the Table is approximately 66~kg, corresponding to a self-load (dead load) of about 660~N, which is represented by red arrows in the Autodesk Inventor simulation, as shown in Figure~\ref{fig:weight_load}. The total mass of the Cage, including the 20 supermodules, is approximately 450~kg, exerting a load on the top inner frames of the table as intended by the design. This load is depicted by yellow arrows in Figure~\ref{fig:weight_load}. Although the actual load is concentrated at both ends of the frame, a uniform pressure load was applied in the simulation to represent a more conservative scenario and to simplify the initial conditions. To account for a potential detector upgrade, an additional safety margin of 2,000~N was included in the evaluation.

\begin{figure}[h]
\begin{center}
  \begin{overpic}[width=0.6\textwidth]{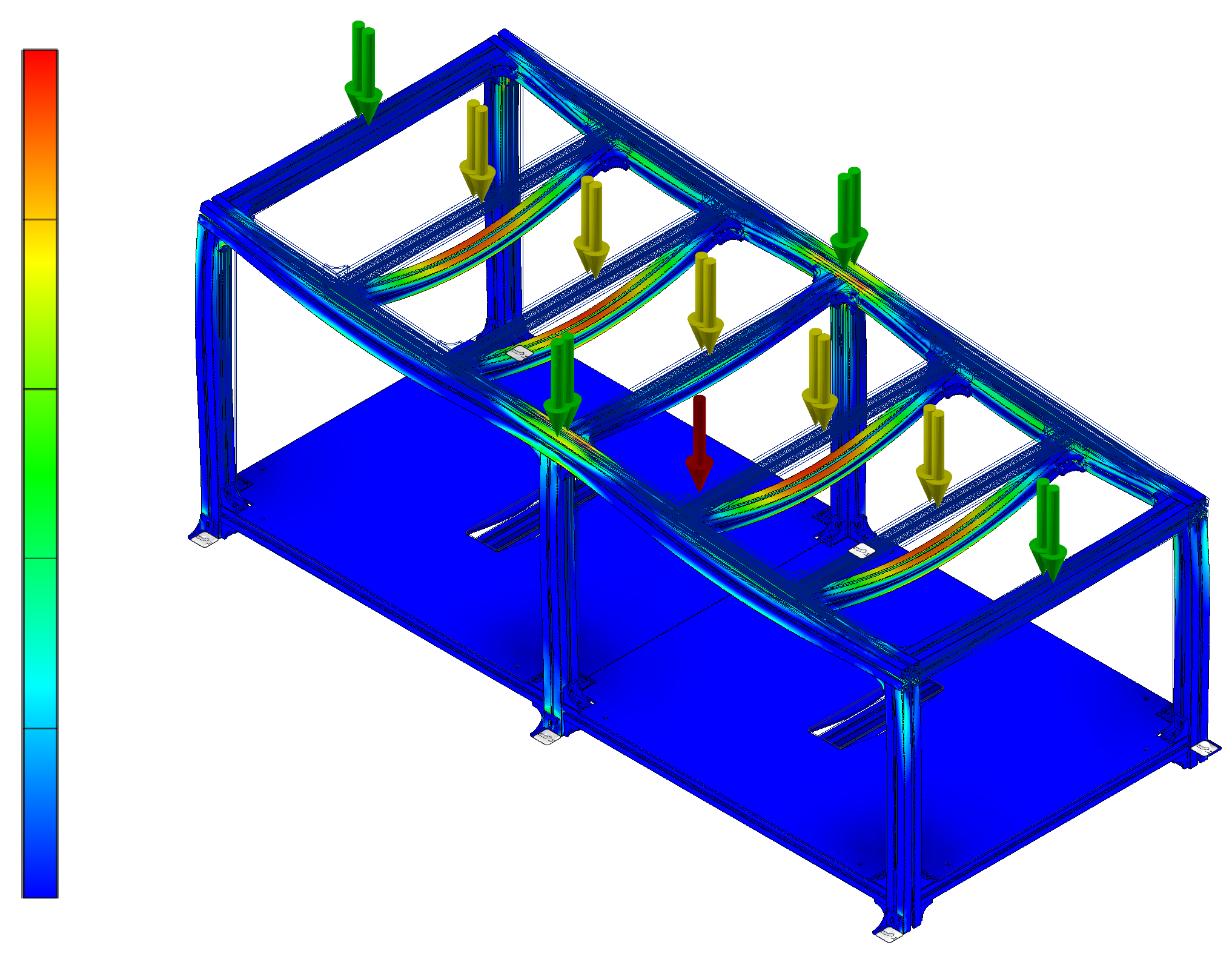}
    \put(13,82){Target side}
    \put(-13,73){15~MPa}
    \put(-13,59){12~MPa}
    \put(-13,46){19~MPa}
    \put(-13,32){ 6~MPa}
    \put(-13,18){ 3~MPa}
    \put(-13, 5){ 0~MPa}
  \end{overpic}
\end{center}
\caption{Von Mises stress results on Table 1 as determined by the \ac{fem} analysis. Color represents the scale of stress.}
\label{fig:vonmises}
\end{figure}
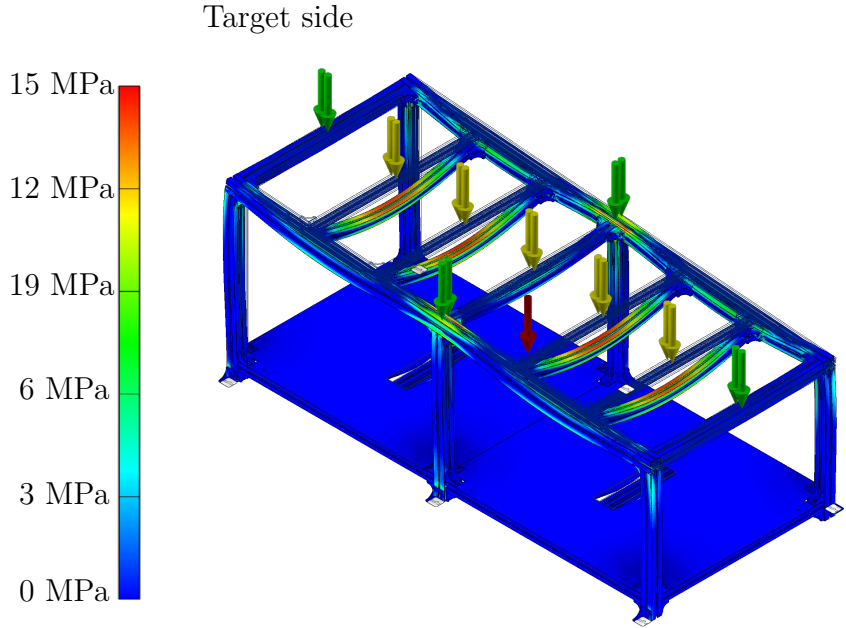

The results of this finite element method (\ac{fem}) analysis include three aspects: the von Mises stress caused by the loads, the safety factor derived from the stress, and the resulting displacement. Figure~\ref{fig:vonmises} shows the von Mises stress induced by both the dead and live loads. According to the analysis, the maximum stress on the Table structure reaches approximately 15~MPa. This stress is primarily concentrated at the center of the horizontal bars, where the load is most intense, while the legs of the Table effectively transfer the load to the ground.

\begin{figure}[h]
\begin{center}
  \begin{overpic}[width=0.6\textwidth]{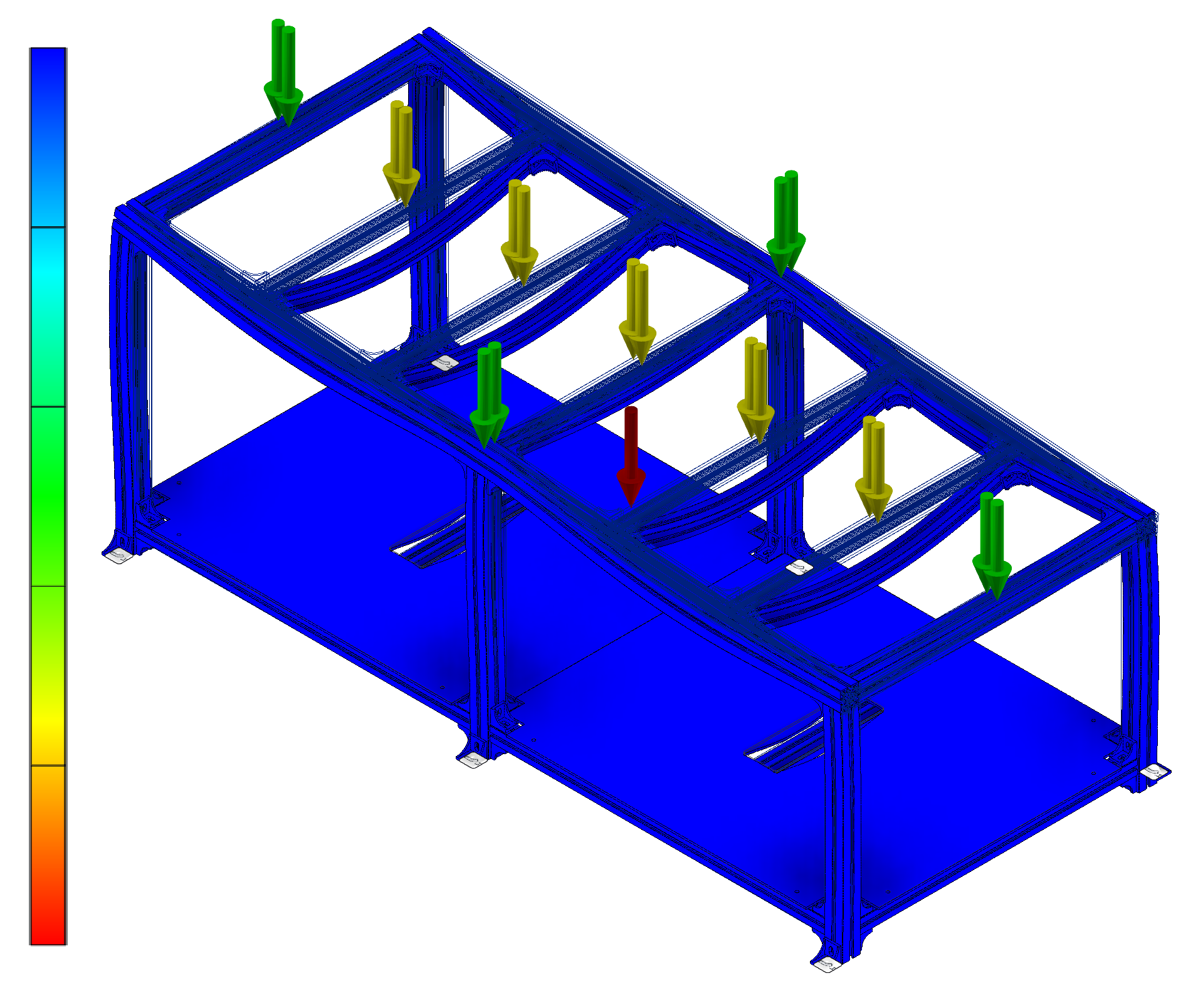}
    \put(10,85){Target side}
    \put(-8,79.5){15.0}
    \put(-8,63.8){14.8}
    \put(-8,48.9){14.6}
    \put(-8,34.1){14.4}
    \put(-8,19.3){14.2}
    \put(-8, 4.5){14.0}
  \end{overpic}
\end{center}
\caption{The safety factor calculated from the stress is 15 in all parts.}
\caption{Safety factor calculated from the stress analysis. It is approximately 15 throughout all components.}
\label{fig:safetyfactor}
\end{figure}

Figure~\ref{fig:safetyfactor} shows the safety factor distribution for each finite element. The safety factor across all parts of the structure is approximately 15, exceeding typical safety requirements~\cite{burr1995mechanical}
\begin{figure}[h]
\begin{center}
  \begin{overpic}[width=0.6\textwidth]{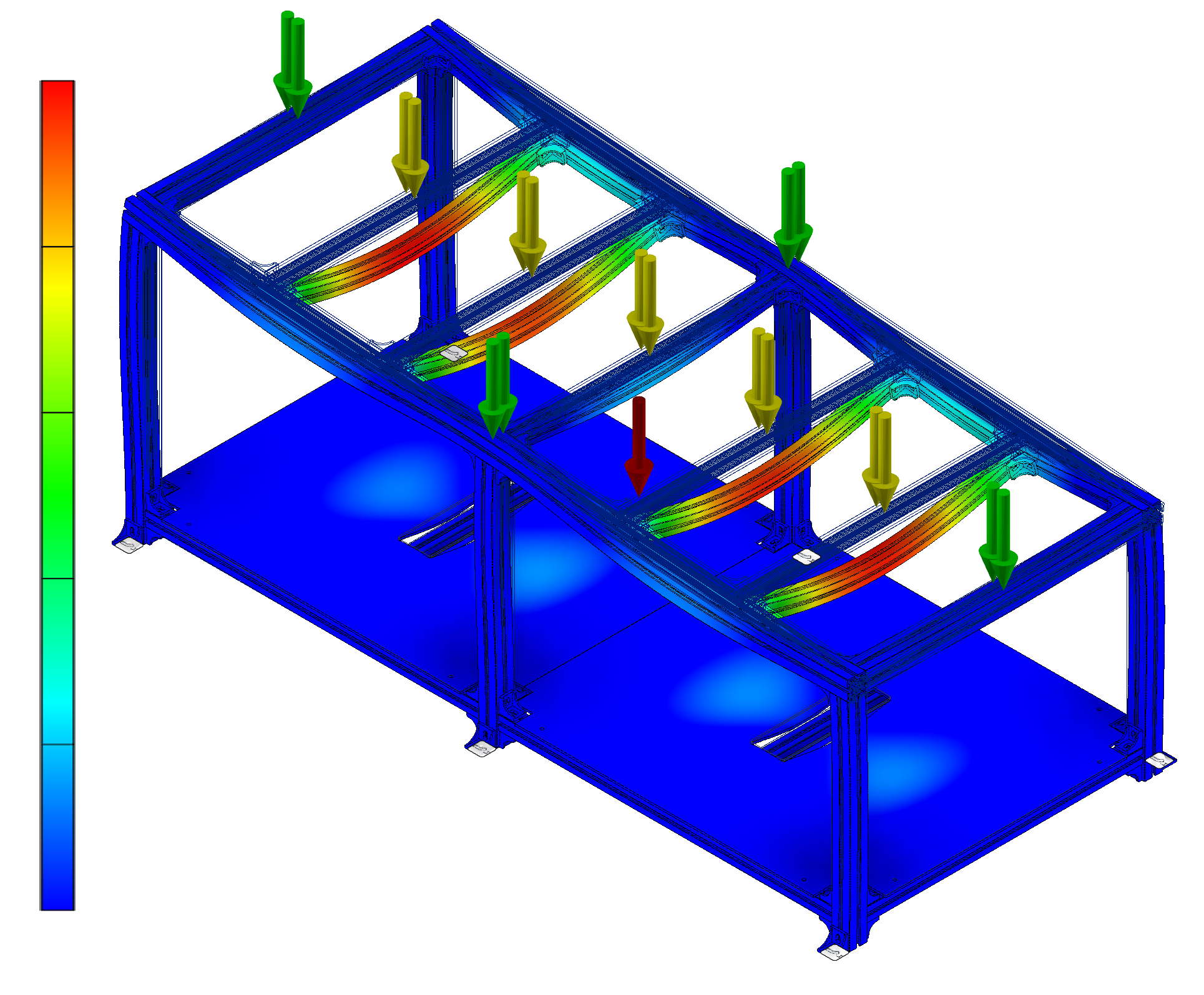}
    \put(10,85){Target side}
    \put(-13,76){1.0~mm}
    \put(-13,62){0.8~mm}
    \put(-13,48){0.6~mm}
    \put(-13,34){0.4~mm}
    \put(-13,20){0.2~mm}
    \put(-13, 6){0.0~mm}
  \end{overpic}
\end{center}
\caption{Results of the weight analysis incorporating a dead load of 660~N force and a live load of 6.5~kN. The maximum displacement observed in the mechanical support segments is 1~mm.}
\label{fig:weight}
\end{figure}
Finally, as shown in Figure~\ref{fig:weight}, the maximum displacement of the mechanical support segments is 1~mm, which is within the spatial tolerance of the supermodules in the Cage. Additionally, since the load of the Cage is supported at both ends rather than at the center of the inner frame, the resulting deformation remains below 1~mm.

The load analysis described above pertains to the simulation of Table 1, which is closer to the target. Table 2 shares the same design as Table 1, except for a reduced height by 212~mm. Due to the reduced height, the center of mass is lowered, the rigidity of the shortened legs is increased, and the load transfer to the ground is improved. As a result, Table 2 exhibits a similar safety factor and a 3\% decrease in deformation compared to Table 1.

\subsection{Seismic Analysis}\label{sec:seismic}
Given the geographical characteristics of \ac{jparc}, seismic resistance is a critical requirement for the detector. Therefore, in addition to the static load analysis described in Section~\ref{sec:weight}, a seismic analysis was conducted to verify that the structural design remains sufficiently robust under earthquake conditions. The dynamic analysis was performed using the \textit{Autodesk Robot Structural Analysis Professional 2025} program. 

To simulate earthquake conditions, the parameterized seismic response spectrum was employed. This standardized spectrum corresponds to a severe earthquake with a peak ground acceleration of up to $0.65g$, where $g$ denotes the gravitational acceleration. The simulation assumes ground vibrations along the $x$- and $y$-axes, respectively, and calculates the resulting maximum displacement of the structure.

\begin{figure}[h]
\begin{center}
  \begin{overpic}[width=0.6\textwidth]{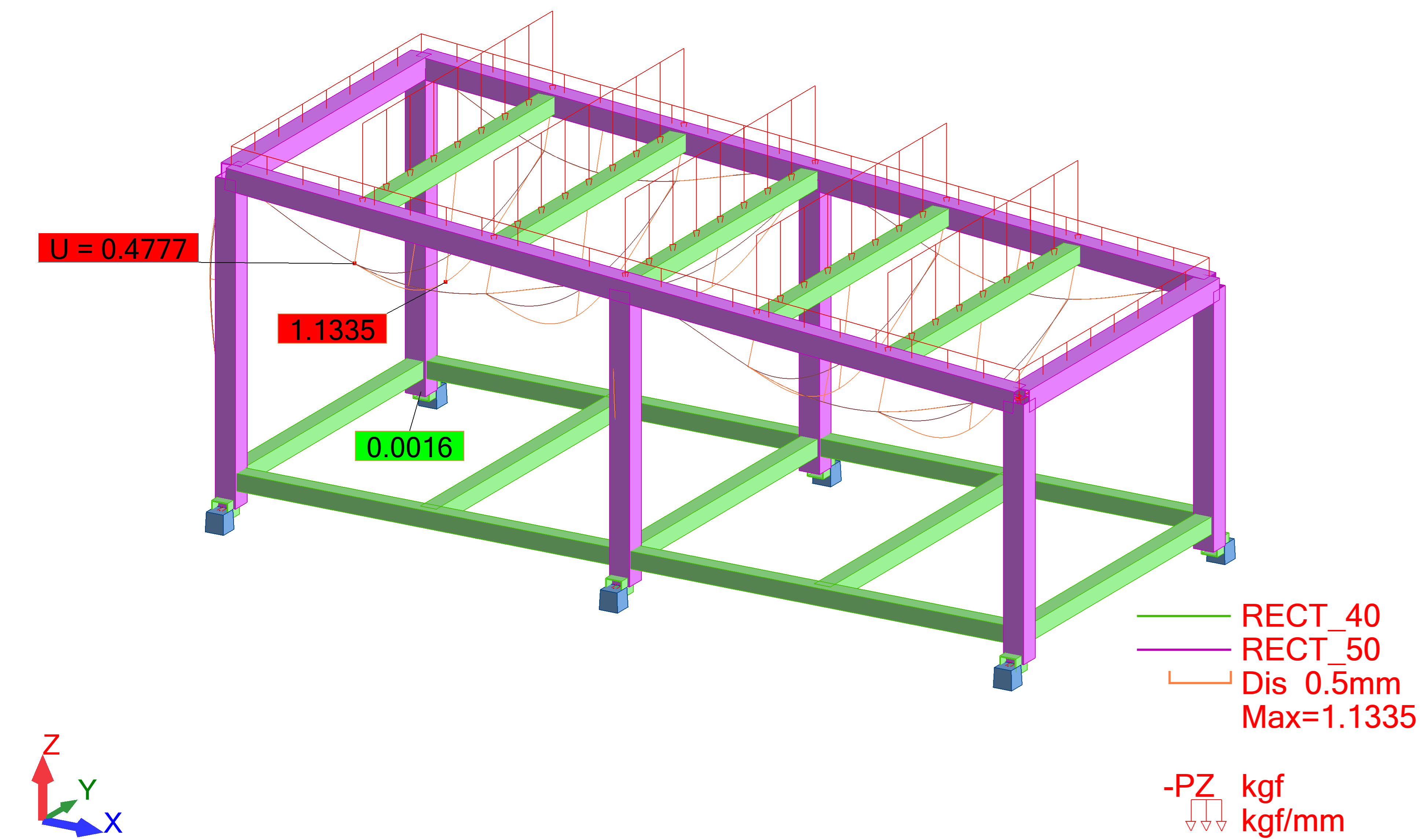}
    \put(-10,30){Target Side}
  \end{overpic}
\end{center}
\caption{Results of the seismic analysis with a dead load of 660~N force, a live load of 6.5~kN, and the seismic spectrum applied in the $y$-direction, corresponding to the worst-case scenario. The maximum displacement of the mechanical support segments is 1.14~mm.}
\label{fig:seismic}
\end{figure}

Figure~\ref{fig:seismic} shows the seismic analysis results in the transverse direction to the detector modules ($y$-direction), representing the worst-case scenario. The maximum displacement of the mechanical support segments is approximately 1.14~mm, which is 0.14~mm greater than in the static load case. 

These results indicate that the detector's response to the design-level ground motion remains within acceptable limits. Furthermore, the support structures and connections were designed to provide sufficient strength and ductility, ensuring stability and functionality during and after seismic events.

The results above pertain to Table~1. Simulation results show that the maximum displacement of Table~2 is also approximately 1.08~mm, indicating that Table~2 is less affected by the earthquake due to its lower height.

In summary, the weight and seismic analyses demonstrate that the \ac{submet} detector is structurally sound and capable of withstanding the anticipated loads throughout its operation. These analyses provide confidence in the long-term stability and safety of the experiment.

\newpage
\section{Summary}
\ac{submet} investigates the existence of millicharged particles, a potential dark matter candidate, within the largely unexplored parameter space of mass $m_\chi < 1.6\,\textrm{GeV}/c^2$ and charge $Q < 10^{-3}e$.

Plastic scintillator is employed as the active material, where millicharged particles produce photons via ionization. To enhance the detection probability, 1.5~m-long scintillator bars are aligned along the expected momentum direction of the millicharged particles. The detector consists of two layers of scintillator stacks, increasing the active volume while effectively controlling backgrounds. Each scintillator bar is coupled to a \ac{pmt}, forming the fundamental detection unit referred to as a module.

In designing the detector, mechanical stability, precise alignment with the beam direction, and ease of access to components post-installation were carefully considered. Modules are grouped in $2 \times 2$ arrays to form supermodules, which are then mounted inside Cages to maintain uniform spacing between modules. The overall design ensures mechanical robustness and seismic resistance, thereby guaranteeing the safe operation of the experiment. The fully assembled detector was successfully installed at the detector site and began data-taking operations in June 2024.

\section*{Acknowledgment}
This work was supported by the National Research Foundation of Korea (NRF) grants funded by the Korea government (MSIT) (RS-2021-NR059935 and RS-2025-00560964) and a Korea University grant.

\vspace{0.2cm}
\noindent
\let\doi\relax
\bibliographystyle{ptephy}
\bibliography{submet_mechanics_ptep}

\end{document}